\documentclass[submission, Phys]{SciPost}
\usepackage{amssymb}
\usepackage{amsmath}
\usepackage{braket}
\usepackage{color}
\usepackage{MnSymbol}
\usepackage[amsmath]{empheq}
\usepackage{tikz}
\usepackage{color}
\usepackage{enumerate}
\usepackage{comment}
\usepackage{nicefrac}
\usepackage{caption}
\usepackage{bbm}
\usepackage{tensor}
\usepackage{cite}
\usepackage{mathtools}

\usepackage{graphicx}
\usepackage{subcaption}
\usepackage{tikz}

\newcommand{\be}{\begin{equation}}
\newcommand{\ee}{\end{equation}}
\newcommand{\ba}{\begin{aligned}}
\newcommand{\ea}{\end{aligned}}

\usepackage{cancel}

\begin{document}

%\title{
%Measurement Catastrophe and  Emergence of Schr\"odinger Cat States
%}
%\author{Lenart Zadnik}
%\affiliation{Universit\'e Paris-Saclay, CNRS, LPTMS, 91405, Orsay, France}
%\author{Saverio Bocini}
%\affiliation{Universit\'e Paris-Saclay, CNRS, LPTMS, 91405, Orsay, France}
%\author{Kemal Bidzhiev}
%\affiliation{Universit\'e Paris-Saclay, CNRS, LPTMS, 91405, Orsay, France}
%\author{Maurizio Fagotti}
%\affiliation{Universit\'e Paris-Saclay, CNRS, LPTMS, 91405, Orsay, France}

%\date{\today}
%
%\begin{abstract}
%In  many-body quantum systems with Hilbert-space fragmentation it is possible to find stationary states manifesting quantum jamming. 
%It was recently shown that these are ``states with memory'', where, \emph{e.g.}, measuring a localised observable has everlasting macroscopic effects. We study such a  measurement catastrophe with the help of an example that stands out for its  clarity. 
%We detail two extraordinary behaviours: (1) The expectation value of a charge density has a nontrivial profile in the ballistic scaling limit (the distance of the observable from the site of the measurement is proportional to the infinitely large time), while the corresponding current approaches zero; (2) At late times, after an extra projective measurement in a specific sub-chain, there is a finite probability for the system to approach a Greenberger-Horne-Zeilinger state with, ideally, arbitrarily many spins.
%\end{abstract}
%\maketitle

\begin{center}{\large \textbf{
Measurement catastrophe and ballistic spread of charge density with vanishing current
}}\end{center}

\begin{center}
L. Zadnik\textsuperscript{1}, S. Bocini\textsuperscript{1}, K. Bidzhiev\textsuperscript{1}, and M. Fagotti\textsuperscript{1}
\end{center}

\begin{center}
{\bf 1} Universit\'e Paris-Saclay, CNRS, LPTMS, 91405, Orsay, France
\\
* maurizio.fagotti@universite-paris-saclay.fr
\end{center}

\begin{center}
\today
\end{center}

\section*{Abstract}
{\bf
One of the features of many-body quantum systems with Hilbert-space fragmentation are stationary states manifesting quantum jamming. It was recently shown that these are ``states with memory'', in which, e.g., measuring a localised observable has everlasting macroscopic effects. We study such a  measurement catastrophe with an example that stands out for its  clarity. We show in particular that at late times the expectation value of a charge density becomes a nontrivial function of the ratio between distance and time notwithstanding the corresponding current approaching zero.
}

\vspace{10pt}
\noindent\rule{\textwidth}{1pt}
\tableofcontents\thispagestyle{fancy}
\noindent\rule{\textwidth}{1pt}
\vspace{10pt}

%\tableofcontents

\section{Introduction}
Complexity is behind the most intriguing phenomena observed in many-body systems: phases of matter, spontaneous symmetry breaking, relaxation, etc.  %On the other hand, our personal understanding of a phenomenon can be either intuitive or, almost inevitably, reductive, being based on simplifications and analogies. 
On the other hand, our personal understanding of a phenomenon is almost inevitably reductive, being based on simplifications and analogies.
Undressing a problem of the complicated details that do not affect the behaviour under investigation is one of the most important steps towards the development of a decent physical picture. Just as the framework of renormalisation group allows one to identify what is relevant at a phase transition, so the conception of a simple model 
helps identify what is essential. 

Exactly solvable models are arguably the main source of procedural knowledge in physics: by solving a problem exactly, we can grasp the significance of what could be seen in nature. 
These models, which are generally integrable, are clearly special. As such, they exhibit exceptional features that could be absent in real systems. A phenomenon identified in an exactly solvable model is nevertheless always potentially observable, if not exactly like predicted in the model, as a somehow related phenomenon. Pre-thermalisation is a remarkable example of this kind: integrable systems are known to relax differently than generic ones~\cite{Kinoshita2006,rigol2007,Cazalilla2006,Eisert2015,Polkovnikov2011}, and yet such special, atypical dynamics leave a mark also on the relaxation of generic systems~\cite{Bertini2015}, which can exhibit pre-thermalisation plateaux~\cite{Langen2016}. 

In fact, even working in the framework of integrability can present formidable complications that undermine our qualitative understanding. This is one of the motivations behind the recent interest in identifying the ``simplest'' interacting integrable systems. In this respect, a spin-$1/2$ chain model with a three-site interaction termed the ``dual folded XXZ model'' was recently highlighted~\cite{folded:1,folded:2,pozsgay2021}. It corresponds to a special point of a rather complicated integrable system called Bariev model~\cite{Bariev1991,Bariev1992}, and is equivalent to the celebrated Heisenberg XXZ model in the limit of large anisotropy. The dual folded XXZ model was recently solved by two of us from scratch by means of a coordinate Bethe Ansatz that has been adapted to the special symmetries emerging in the large-anisotropy limit of the XXZ model~\cite{folded:1}. In particular, the Hilbert space splits into an exponentially large number of sectors with a given configuration of particles, each particle being associated with a spin up in either an even or an odd position. Similar structures arise in what was recently dubbed ``Hilbert-space fragmentation''~\cite{sala2020,khemani2020,moudgalya2021}.
In each sector the model is almost free, with a diagonal scattering matrix in which only one diagonal element is different from $-1$ and equal to  $-e^{i q}$, where $q$ is the difference between the momenta of the scattering particles, in a remarkable analogy with the scattering phase modification factor in hard-rod deformed models (see, e.g., Refs~\cite{pozsgay2021hardrod,cardy2021toverline}). A striking feature found in fragmented models, such as the folded XXZ one, is the existence of an exponentially large sector consisting of jammed states: particles are stuck and can not move~\cite{moudgalyaKrylov2021,HSF,Turner2018,katja2020,bastianello2021fragmentation,Menon1997}. In Ref.~\cite{measurement2021} it was shown that such states have memory: the effect of a localised perturbation does not fade away but rather becomes visible at macroscopic scales for arbitrarily long times. 
This is similar to the memory effects observed when perturbing ground states with spontaneously broken symmetries~\cite{Zauner2015,Eisler2020} (see also Refs~\cite{Eisler2020,Eisler2016,Gruber2021,Eisler2021} for a perturbation local in the fermionic basis), where however the dynamics are effectively constrained in a finite-dimensional space. In the setting of Ref.~\cite{measurement2021}, instead, an exponentially large number of states are involved.
Considering how dramatic the change in the state is, it seems appropriate to refer to it as a ``measurement catastrophe''. In particular, this effect was reported in a class of states that allow for an exact analytic analysis but are, however, not simple enough to make us gain a clear understanding of the phenomenon. This paper is first of all a solution to that problem. 

We consider a family of jammed states for which it is possible to write down the time evolving state after a spin flip in a simple, elegant way. We then focus on observables whose expectation values can be predicted with minimal information about the initial state. Among them there will be the densities of two conserved charges: the magnetisation and the staggered magnetisation. They will be our key monitor of the measurement catastrophe: indeed, they turn out to be affected by the original spin flip in a macroscopic way. This sets the example herein apart from the one considered in Ref.~\cite{measurement2021}, where instead, in the long time limit, charge densities and currents become independent of their positions. The central part of the paper will be devoted to devising a physical picture that could reconcile the nontrivial profile of the (staggered) magnetisation at the Euler/ballistic scale with the fact that the corresponding currents vanish in the limit of infinite time. First we will present a set of rules that allow us to predict the asymptotic behaviour of any spin
solely from the knowledge of the initial configuration of few neighbouring spins.
Using the rules we will then show that the asymptotic state of the system is locally jammed and that the currents of the magnetisation and the staggered magnetisation, as well as all other conventional charges of the folded XXZ model, vanish in it. By computing the statistics of the particles' positions, we will show that the macroscopic reorganisation of the spin configuration coincides with small correlated microscopic fluctuations of particles (spins up) that do not exceed two lattice sites. Finally, we will focus on a special class of weakly interacting initial jammed states, which allow for the preparation of correlated spin pairs in an antiferromagnetic background. We will study, in particular, the entanglement between spins.

\subsection{Overview of the model}

The dual folded XXZ model describes the time evolution of states in the large-anisotropy limit of the Heisenberg model. It is obtained in the ``folded picture'', which was defined in Ref.~\cite{folded:1} as a representation of time evolution with Hamiltonians that have a large coupling constant. Based on the strong coupling expansion of Ref.~\cite{macdonald1988}, the ``folded picture'' refers to a decomposition of time evolution into a periodic high-frequency evolution of operators and a ``gentle'' evolution of the state with a time-independent Hamiltonian.
The dual folded XXZ Hamiltonian reads
\begin{align}
\label{eq:Hamiltonian}
    H=\mathcal{J}\sum_{\ell}\frac{1- \sigma_{\ell+1}^z}{2}(\sigma_\ell^x \sigma_{\ell+2}^x+ \sigma_\ell^y \sigma_{\ell+2}^y)\, .
\end{align}
It is integrable and describes constrained hopping on a lattice, summarised by the following local dynamical rule: $\ket{\uparrow\downarrow\downarrow}\,\leftrightarrow\,\ket{\downarrow\downarrow\uparrow}$. Such a rule preserves the sequence of parities of the positions of spins up. This in turn implies conservation of the configuration of two species of particles defined by that parity. The presence of multiple species of particles typically signals that an integrable model is solvable via a complicated nested Bethe Ansatz. And indeed, the dual folded XXZ model corresponds to a strong-repulsion regime of the two-component Bariev model~\cite{Bariev1991,Bariev1992}, which is of that type. Nevertheless, at the special point described by Eq.~\eqref{eq:Hamiltonian} the nested structure of the Bariev model is reduced to such a degree that Bethe equations can be solved explicitly. Perhaps the most striking remnant of the nested structure is the conservation of the configuration, which results in the breakdown of one-site shift invariance into two-site shift invariance.

The constrained hopping leads to a very rich Hilbert-space structure combined with an exceptional simplicity of the Bethe equations. It enables exact analytical and numerical analysis of various phenomena that have recently been in the spotlight: Hilbert-space fragmentation~\cite{HSF}, pre-relaxation~\cite{Fagotti2014}, generalised hydrodynamics~\cite{folded:2}, spin transport phenomena~\cite{sarang2021}, the effect of non-abelian symmetries, time-translation symmetry breaking~\cite{pozsgay2021}, and the macroscopic effects of local measurements~\cite{measurement2021}. Most of these phenomena are accessible in sectors that allow for an even simpler description; we mention, for example, the non-interacting sectors, where only one species of particles is present, or the exponentially large jammed sector, where nearest-neighbour pairs of spins down are prohibited and the basis states can be chosen to be eigenstates of all $\sigma^z_\ell$ simultaneously. 

The local conserved charges of the dual folded XXZ model have been classified in Ref.~\cite{folded:1}. Their place in the Yang-Baxter integrability structure of the model~\cite{zhou96,shiroishi97,Zhang2020} was described in Refs.~\cite{balazs2021a,balazs2021b}. We mention that the fragmentation of the Hilbert space in the folded XXZ model could be related to the existence of additional non-abelian conservation laws, potentially associated with the emergence of the integrable analogue (cf. Refs~\cite{zadnik16,medenjak20,miao21}) of quantum many-body scars~\cite{moudgalya2021,vafek2017,Beri2019,sanjay2021,sanjay2020,mark2020}.

\section{Initial state and spin-flip protocol}\label{s:protocol}

We consider a protocol that consists of the following steps:
\begin{enumerate}
\item Prepare the system in a \emph{jammed} product state with spins aligned along the $z$-axis (in either direction).

In a jammed state fragments of the form $\cdots\downarrow\downarrow\cdots$ are prohibited. We use the notations of Ref.~\cite{folded:1}: spins up correspond to particles of two species determined by the parity of their positions. Specifically, a spin up at site $\ell$ corresponds to a particle of species $b=2\lceil\ell/2\rceil-\ell\in\{0,1\}$ at macrosite $\ell'=\lceil\ell/2\rceil$. A jammed state is completely characterised by the configuration $\underline b=(\ldots, b_n,b_{n+1},\ldots)$ of spins up as follows:
\begin{align}
\ket{\underline b}=\prod_{j}\sigma_{2\ell_j'-b_j}^+\ket{\downarrow\cdots\downarrow}\, .
\end{align}
Here, the macrosites satisfy the recurrence relation
\begin{align}
\label{eq:recurrence}
\ell'_{j+1}=\ell'_{j}+1-b_{j}(1-b_{j+1})\, ,
\end{align}
according to which two subsequent particles of species $1$ and $0$, respectively, occupy the same macrosite. This prevents state fragments of the form $\cdots\underline{\uparrow\downarrow}\,\underline{\downarrow\uparrow}\cdots$ (underlined are the macrosites), which violate the jamming condition. 

\item Flip a spin up that was close to a spin down (otherwise the state remains jammed). This step can be interpreted as a result of a double projective measurement, e.g., measuring $\sigma^z$ after $\sigma^x$.

The spin flip destroys a particle. We choose its index to be $0$, so that the state now reads
\be
\Bigl(\prod_{j\le-1}\sigma^+_{2\ell'_j-b_j}\Bigr)\Bigl(\prod_{j\ge 1}\sigma^+_{2\ell'_j-b_j}\Bigr)\ket{\downarrow\cdots\downarrow}\, .
\ee
Recurrence~\eqref{eq:recurrence} can now be solved with respect to the position of the spin flipped. First we note that either $b_{-1}$ or $b_1$ should be equal to $b_0$: we flipped a spin close to a spin down (the site of the next spin up has the same parity as the site of the spin flipped). We choose the macrosites in such a way that at $\ell'_0=0$ there are two spins down. As a result we either have $b_{-1}=b_0=0$ or $b_{1}=b_0=1$.
In fact, since we removed a particle (a spin up), it is convenient to redefine the index of the particles, $j$, so as to fill the hole at $j=0$. We choose the convention $j_{\rm new}=j_{\rm old}+\theta(j_{\rm old}<0)$ and then find
\be\label{eq:mapping0}
\ell'_j=j-\theta(j\leq 0)+\begin{cases}
-\sum^{j-1}_{m=1}b^{(0)}_{m}(1-b^{(0)}_{m+1})\,, &\text{if }j>0\\
\sum^{|j|}_{m=1}(1-b_{1-m}^{(0)}) b_{-m}^{(0)}\,, &\text{if }j\leq 0\, .
\end{cases}
\ee
Here, $\underline b^{(0)}$ is the configuration obtained by removing $b_0$ from $\underline b$: $b^{(0)}_j=b_{j-\theta(j\leq  0)}$.
We indicate this state by 
\be
\ket{\Psi(t=0)}=\ket{0;\underline b^{(0)}}\, ,
\ee
where $\ket{n;\underline c}$ denotes a state which is jammed everywhere except between the particles $c_n$ and $c_{n+1}$, where either {\em two} or {\em three} consecutive spins down are present. We call this excess of spins down ``impurity''. 
\item Time evolve with $H$.

While the folded XXZ model is interacting, time evolution is particularly simple in the sector of our initial state. As shown in Appendix~\ref{app:time_evolution_of_state}, this sector is spanned by states $\ket{n;\underline b^{(0)}}$, in which the particle macrosites read 
\begin{align}
\label{eq:mapping1}
\ell^{\prime(n)}_j=j-\theta(j\leq n)+\begin{cases}
-\sum^{j-1}_{m=1}b^{(0)}_{m}(1-b^{(0)}_{m+1})\,, &\text{if }j>0\\
\sum^{-j}_{m=1}(1-b_{1-m}^{(0)}) b_{-m}^{(0)}\,, &\text{if }j\leq 0\, .
\end{cases}
\end{align}
They are obtained from Eq.~\eqref{eq:mapping0} by noting that, when the impurity jumps across a particle, the macrosite of the latter shifts by $1$ in the opposite direction.
\end{enumerate}

\section{Exact solution to the dynamics}

The configuration of particles in the state is preserved under the action of the Hamiltonian~\eqref{eq:Hamiltonian}. As a result, the impurity can be interpreted as a fermion freely hopping in the background $\underline{b}^{(0)}$ of particles. As shown in Appendix~\ref{app:time_evolution_of_state}, the state at time $t$ reads
\be
\ket{\Psi(t)}=\sum_n (-i)^n J_n(4\mathcal{J} t)\ket{n;\underline b^{(0)}}\, ,
\ee
where $J_n(x)$ are Bessel functions.

Let us denote the expectation value $\bra{\Psi(t)}A\ket{\Psi(t)}$ by $\braket{A}_t$ for any operator $A$ at arbitrary time $t$. For the expectation values of operators $D$ that are diagonal in the basis of states $\ket{n;\underline{b}^{(0)}}$, for example $D=\sigma_\ell^z$, we now have
\be
\label{eq:diagonal_expectation}
\braket{D}_t=\sum_n  |J_n(4\mathcal{J} t)|^2\braket{n;\underline b^{(0)}|D|n;\underline b^{(0)}}\, .
\ee
If $D^2=1$, like for a string of Pauli matrices $\sigma_\ell^z$, the expectation values $\braket{n;\underline b^{(0)}|D|n;\underline b^{(0)}}$ can be either $1$ or $-1$. 

The subtle difference between these dynamics and the ones in the XX model, in which the initial state has a single spin up, is hidden in the mapping~\eqref{eq:mapping1} between the positions in the chain and those in the configuration. As will be shown in the following, this mapping has striking effects on the dynamics of local observables,  such as the $z$-component of the spin.

\subsection{Macrosite fluctuations of particles}
\label{sec:fluctuations}

We embrace here the Lagrangian perspective of following the motion of particles. A particle is characterised by its species and its macro-position. The species is fixed whereas the macro-position is a dynamical variable. In the following we compute the  statistics of the particles' macro-positions. To that aim, let us consider the macrosite~\eqref{eq:mapping1} corresponding to the particle labeled by $j$ (i.e., the $j$-th spin up). This can be thought of as the eigenvalue of an operator that is diagonal in the basis of states $\ket{n;\underline{b}^{(0)}}$, in which it assumes values $\theta(j>n)+c(j)$, where $c(j)$ denotes is its state-independent part.
%\footnote{From now on we will suppress the upper index $n$ referring to the state, i.e., we will denote the macrosite simply by $\ell_j'$.}
Using Eq.~\eqref{eq:diagonal_expectation} we can now compute the  characteristic function $\braket{e^{i k \ell'_j}}_t$. We find
\be
\label{eq:characteristic_fn}
\braket{e^{i k \ell'_j}}_t=e^{i k c(j)}[(e^{i k}-1)f(j,t)+1]\, ,
\ee
where we introduced the auxiliary quantity
\begin{align}
\label{eq:f_function}
    f(x,t):=\sum_n \theta(x>n)|J_n(4\mathcal{J}t)|^2\sim\frac{1}{2}+\frac{1}{\pi}\arcsin\Big(\frac{x}{4\mathcal{J} t}\Big)\, .
\end{align}
The characteristic function~\eqref{eq:characteristic_fn}, in particular, implies
\begin{align}
\mathrm{Prob}(\ell'_j)=\begin{cases}
1-f(j,t)\,,&\text{if }\ell'_j=c(j)\\
f(j,t)\,,&\text{if }\ell'_j=c(j)+1\, ,
\end{cases}
\end{align}
as well as
\begin{align}
    \braket{\ell'_j}_t=f(j,t)+c(j)\,,\qquad \braket{\ell_j'^2}_t^{\rm c}=f(j,t)-f(j,t)^2\,,
\end{align}
where $\langle A B\rangle^{\rm c}_t:=\langle AB\rangle_t-\langle A\rangle_t\langle B\rangle_t$ denotes the connected correlation function. 

\begin{figure}[ht!]
\centering
\hspace{-1.75em}\includegraphics[width=0.5\textwidth]{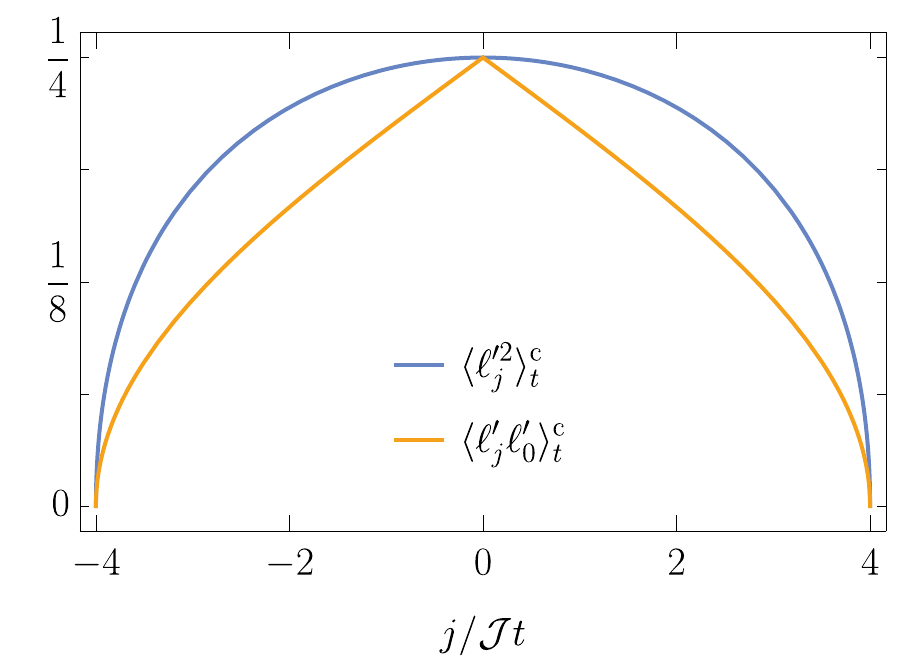}
\caption{Fluctuations of the macrosite corresponding to the $j$-th particle, and the equal-time connected correlation function between the macrosites of particle $0$ and particle $j$. Both are plotted as functions of the ballistic ray $j/(\mathcal{J}t)$.}
\label{fig:fluctuations}
\end{figure}

The particles therefore experience fluctuations of order $O(1)$ around their average positions. The fluctuations of different particles are moreover coherent, as can be inferred from the following non-negative equal-time connected correlation function
\begin{align}
    \braket{\ell'_m\ell'_n}_t^{\rm c}=\theta(m\ge n) f(n,t)+\theta(n>m)f(m,t)-f(m,t)f(n,t)\, ,
\end{align}
plotted along with $\braket{\ell_j'^2}_t^{\rm c}$ in Fig.~\ref{fig:fluctuations}.

\subsection{Local magnetisation}
\begin{figure}[t!]
  \hspace{-1.em}
  \centering
  \includegraphics[width=0.95\textwidth]{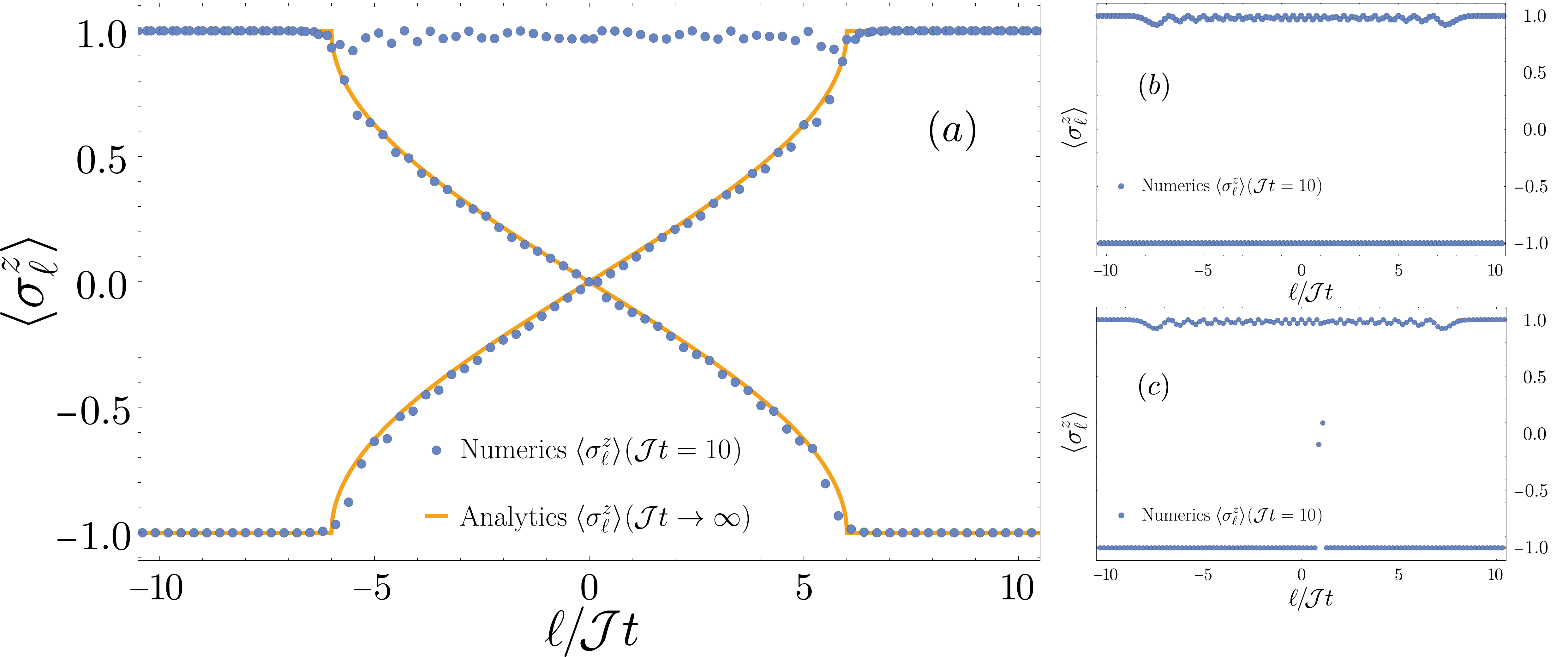}
\caption{Panel (a) shows $\braket{\sigma^z_\ell}_t$ as a function of the ballistic ray coordinate $\ell/(\mathcal{J}t)$. The initial state (immediately after the spin flip) is $\ket{\cdots\uparrow\uparrow\downarrow\uparrow\uparrow\downarrow\uparrow \downarrow \downarrow\uparrow\uparrow\downarrow\uparrow\uparrow\downarrow\cdots}$. Blue points correspond both to numerical implementation of Eq.~\eqref{eq:diagonal_expectation} with $D=\sigma_\ell^z$, as well as to the DMRG data (they are in perfect agreement). The orange line corresponds to the asymptotic behaviour~\eqref{eq:asymptotic_magnetisation}, in which $x(\ell)=2\ell/3$, as follows from Eq.~\eqref{eq:x_to_ell}. 
Panel (b) shows $\braket{\sigma^z_{\ell}}_t$ for $\mathcal{J}t=10$ for a state with one species of particles, i.e., a noninteracting state. For the initial configuration we chose a N\'eel state with one of the spins up flipped down $\ket{\cdots\uparrow\downarrow \uparrow\downarrow \downarrow\downarrow \uparrow\downarrow \uparrow\downarrow\cdots}$. Panel (c) shows $\braket{\sigma^z_{\ell}}_t$ for $\mathcal{J}t=10$ in the state given in Panel (b) with complementary flipped up spin at $\ell = 10$. Two blue points describe the spins not aligned with the $z$-axis.
}
  \label{fig:Folded_magnetization}
\end{figure}

We now adopt an Eulerian point of view and study the properties of the state at a given position or ray.
The Hamiltonian commutes with $\prod_j\sigma_j^z$ and the initial state is an eigenstate of that operator, therefore $\braket{\sigma_\ell^x}_t=\braket{\sigma_\ell^y}_t=0$. The local magnetisation along the $z$-axis instead exhibits a nontrivial ballistic-scale profile. Indeed, as shown in Appendices~\ref{app:matrix_elements} and \ref{app:profiles_magnetisation}, the asymptotic behaviour of the local magnetisation along the $z$-axis is determined by the following rules:
\begin{align}
\label{eq:asymptotic_magnetisation}
\braket{\sigma_{\ell}^z}_t\sim \begin{cases}
1\,,&\text{for }\ket{\cdots \bullet\cdots\uparrow_\ell\updownarrow\uparrow\cdots}\vee \ket{\cdots\uparrow\updownarrow\uparrow_\ell\cdots\bullet\cdots}\\
\frac{2}{\pi}\arcsin\big(\tfrac{x(\ell)}{4\mathcal{J} t}\big)\,,&\text{for }\ket{\cdots\bullet\cdots\uparrow_\ell\uparrow\downarrow\cdots}\vee \ket{\cdots\uparrow\uparrow\downarrow_\ell\cdots\bullet\cdots}\\
-\frac{2}{\pi}\arcsin\big(\tfrac{x(\ell)}{4\mathcal{J}t}\big)\,,&\text{for }\ket{\cdots\bullet\cdots\downarrow_\ell\uparrow\uparrow\cdots}\vee \ket{\cdots\downarrow\uparrow\uparrow_\ell\cdots\bullet\cdots}\\
-1\,,&\text{for }\ket{\cdots\bullet\cdots\downarrow_\ell\uparrow\downarrow\cdots}\vee\ket{\cdots\downarrow\uparrow\downarrow_\ell\cdots\bullet\cdots}\,.
\end{cases}
\end{align}
Here, $\ket{\cdots\bullet\cdots\updownarrow_\ell\cdots}$ and $\ket{\cdots\updownarrow_\ell\cdots\bullet\cdots}$ stand for fragments of the initial state in which $\ell$ is positive or negative, respectively. For large-enough $|\ell|$, $x(\ell)$ can be computed from 
\begin{align}
    \label{eq:x_to_ell}
    \frac{1}{\ell}\int_0^{x(\ell)}\mathrm d n\, \xi(n)=\frac{1}{2}\, ,
\end{align}
where $\xi(n)$ is the coarse-grained macro-distance between particles, namely the average of $1-b^{(0)}_j(1-b^{(0)}_{j+1})$ over a large-enough interval of $j$ including %the particle with index 
$n$. For example, $\xi(n)=3/4$ for the fragment $\ldots,1,1,0,0,1,1,0,0,1,1,0,0,\ldots$ The remarkable feature of Eq.~\eqref{eq:asymptotic_magnetisation} is that the asymptotic value of the local magnetisation at site $\ell$ is determined solely by a configuration of three consecutive spins around that site. The only sites in which the spins are asymptotically not aligned along the $z$-axis are those that start in one of the following spin configurations: $\bullet\cdots\underline{\uparrow}\!\uparrow\downarrow$,  $\bullet\cdots\underline{\downarrow}\!\uparrow\uparrow$, $\uparrow\uparrow\!\underline{\downarrow}\cdots\bullet$, and  $\downarrow\uparrow\!\underline{\uparrow}\cdots\bullet$. Their common feature is the existence of both species of particles and therefore the presence of interaction~\cite{folded:1}. Indeed, as shown in Fig.~\ref{fig:Folded_magnetization}, initial states with a single species of particles do not exhibit nontrivial ballistic profiles.

%%%
%%%

\subsection{Emergence of a locally quasi-jammed state}
Since the spin flip at time $t=0$ creates an excess of spins down, for $t>0$ the state is not jammed anymore. However, the weaker condition
\begin{align}\label{eq:LQJcond}
P_{\downarrow\downarrow}(\ell)=\bigl\langle\frac{1-\sigma_\ell^z}{2}\frac{1-\sigma_{\ell+1}^z}{2}\bigr\rangle_t\xrightarrow{\mathcal{J}t\rightarrow\infty}0\, ,\quad \text{for all $\ell$}\, ,
\end{align}
used in Ref.~\cite{measurement2021} to define locally quasi-jammed states (LQJS), still holds. Expressed in the basis of states $\ket{n;\underline{b}^{(0)}}$, the expectation value used in the LQJS condition reads (see Appendix~\ref{app:correlators})
\begin{align}
\label{eq:jamming2}
    P_{\downarrow\downarrow}(\ell)=\sum_n|J_n(4\mathcal{J}t)|^2\braket{n;\underline{b}^{(0)}|\frac{1-\sigma_{\ell}^z}{2}|n;\underline{b}^{(0)}}\braket{n;\underline{b}^{(0)}|\frac{1-\sigma_{\ell+1}^z}{2}|n;\underline{b}^{(0)}}\, .
\end{align}

\begin{figure}[t!]
  \hspace{-1.5em}\centering
  \includegraphics[width=0.6\textwidth]{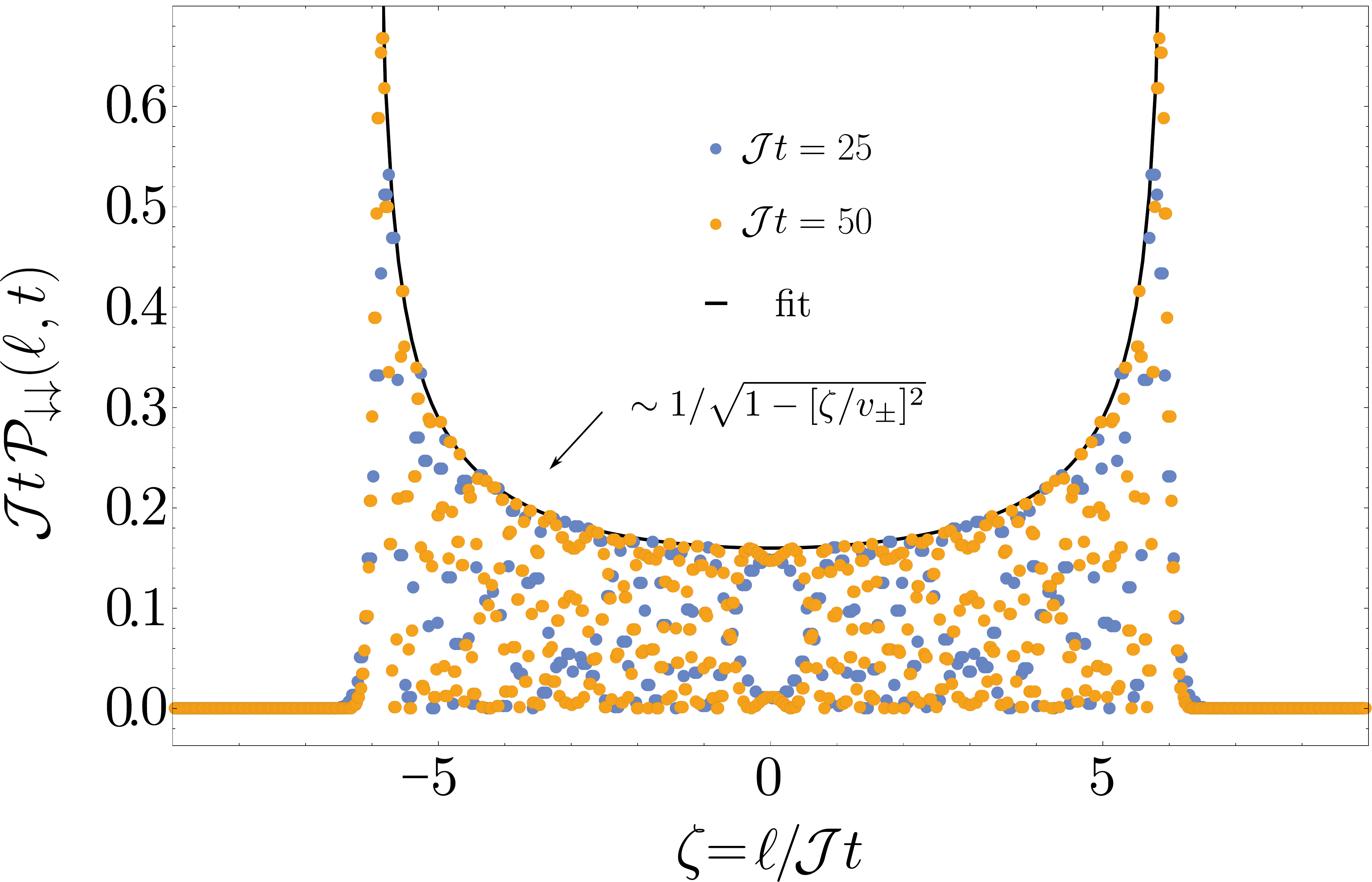}
  \caption{Rescaled jamming condition profile $\mathcal{J}t\mathcal{P}_{\downarrow\downarrow}(\ell,t)$ for $\mathcal{J}t=25$ (blue) and $\mathcal{J}t=50$ (orange) in the initial state  $\ket{\cdots\uparrow\uparrow\downarrow\uparrow\uparrow\downarrow\uparrow \downarrow \downarrow\uparrow\uparrow\downarrow\uparrow\uparrow\downarrow\cdots}$. The DMRG data confirm the $O(1/t)$ approach towards an LQJS: $\mathcal{J}tP_{\downarrow\downarrow}(\ell,t)$ forms a profile with a scale invariant envelope, conjectured to be $a/\sqrt{1-[\zeta/v_\pm]^2}$, where $a\approx 0.16$, $v_\pm = \pm 8\mathcal{J}/\left(1+\tfrac{2}{L}\braket{S^z}_0\right) = \pm 6$ ($\braket{S^z}_0$ is the total magnetisation in the initial state) -- see Ref.~\cite{measurement2021}.
  }
  \label{fig:Folded_Jamming_condition}
\end{figure}

From Eq.~\eqref{eq:jamming2} it follows that the LQJS condition is satisfied up to corrections of order $O(1/t)$, a feature corroborated by exact numerics -- see Fig.~\ref{fig:Folded_Jamming_condition} (an exact analytical calculation in a weakly interacting scenario is presented in Section~\ref{s:special_protocol}). Indeed, 
only few terms of the sum in Eq.~\eqref{eq:jamming2} are different from zero. Let us call $j(\ell)$ the index of the particle represented by a spin up at site $\ell$, or at one of the neighbouring sites if the spin at $\ell$ is down. For $|n-j(\ell)|> n_0$, with $n_0$ a finite positive integer, the matrix elements of $\sigma^z_\ell$ and $\sigma_{\ell+1}^z$ have simple asymptotic forms derived in Appendix~\ref{app:matrix_elements}. Using those, we can compute the eigenvalues of $\tfrac{1}{2}(1-\sigma^z_\ell)$ and
$\tfrac{1}{2}(1-\sigma^z_{\ell+1})$ for all possible configurations of spins in the vicinity of $\ell$: they are shown in Table~\ref{tab:matrix_elements} for a positive $\ell$. Observing that their product is always zero, we conclude that $P_{\downarrow\downarrow}(\ell)$ is a finite sum of squared Bessel functions, which decay to $0$ as $1/t$.
\begin{table}[h!]
\begin{enumerate}[(a)]
\item $n-j(\ell)>n_0$:\\
\begin{tabular}{l || c | c | c | c | c | c | c | c} 
 & $\uparrow_\ell \uparrow \downarrow \uparrow$
 & $ \uparrow_\ell \uparrow  \uparrow \downarrow$
 & $\uparrow_\ell \uparrow  \uparrow \uparrow$  
 & $\uparrow_\ell \downarrow  \uparrow \uparrow$ 
 & $\uparrow_\ell \downarrow  \uparrow \downarrow$
 & $ \downarrow_\ell  \uparrow \uparrow \uparrow$
 &$\downarrow_\ell  \uparrow \uparrow \downarrow$
 & $ \downarrow_\ell  \uparrow \downarrow \uparrow$ \\
 \hline
 $\braket{n;\underline{b}^{(0)}|\tfrac{1}{2}(1\!-\!\sigma_{\ell}^z)|n;\underline{b}^{(0)}}$
 & 1 & 0 & 0 & 0 & 0 & 0 & 0 & 1 \\
 \hline
 $\braket{n;\underline{b}^{(0)}|\tfrac{1}{2}(1\!-\!\sigma_{\ell+1}^z)|n;\underline{b}^{(0)}}$
 & 0 & 1 & 0 & 0 & 1 & 0 & 1 & 0
\end{tabular}

\item $n-j(\ell)<-n_0$:\\
\begin{tabular}{l || c | c | c | c | c | c | c | c} 
 & $\uparrow_\ell \uparrow \downarrow \uparrow$
 & $ \uparrow_\ell \uparrow  \uparrow \downarrow$
 & $\uparrow_\ell \uparrow  \uparrow \uparrow$  
 & $\uparrow_\ell \downarrow  \uparrow \uparrow$ 
 & $\uparrow_\ell \downarrow  \uparrow \downarrow$
 & $ \downarrow_\ell  \uparrow \uparrow \uparrow$
 &$\downarrow_\ell  \uparrow \uparrow \downarrow$
 & $ \downarrow_\ell  \uparrow \downarrow \uparrow$ \\
 \hline
 $\braket{n;\underline{b}^{(0)}|\tfrac{1}{2}(1\!-\!\sigma_{\ell}^z)|n;\underline{b}^{(0)}}$
 & 0 & 0 & 0 & 0 & 0 & 1 & 1 & 1 \\
 \hline
 $\braket{n;\underline{b}^{(0)}|\tfrac{1}{2}(1\!-\!\sigma_{\ell+1}^z)|n;\underline{b}^{(0)}}$
 & 0 & 0 & 0 & 1 & 1 & 0 & 0 & 0
\end{tabular}
\end{enumerate}
\caption{The expectation values of the projectors $\tfrac{1}{2}(1-\sigma_\ell^z)$ and $\tfrac{1}{2}(1-\sigma_{\ell+1}^z)$ in the states $\ket{n;\underline{b}^{(0)}}$ depend only on the initial configuration of the four spins on sites $\ell$, $\ell+1$, $\ell+2$, and $\ell+3$. For $|n-j(\ell)|> n_0$, with $n_0$ a finite positive integer, the products of these expectation values are always zero.}
\label{tab:matrix_elements}
\end{table}

%%%
%%%

\subsection{Local conservation laws and currents}
\label{sec:charges_currents}
The spin along the $z$-axis is related to the macrosite densities of two local conservation laws
\begin{align}
  S^z=\sum_{\ell'}\tfrac{1}{2}(\sigma_{2\ell'-1}^z+\sigma_{2\ell'}^z)\, ,\qquad S_-^z=\sum_{\ell'}\tfrac{1}{2}(\sigma_{2\ell'}^z-\sigma_{2\ell'-1}^z)\, ,
\end{align}
i.e., the magnetisation and the staggered magnetisation, respectively. Using the asymptotic rules~\eqref{eq:asymptotic_magnetisation} we obtain
\be
\Bigl\langle\frac{\sigma_{2\ell'}^z+\sigma_{2\ell'-1}^z}{2}\Bigr\rangle_t\sim\begin{cases}
1\,,&\text{for }\ket{\cdots \bullet\cdots(\uparrow\uparrow)_{\ell'}\uparrow\uparrow\cdots}\vee \ket{\cdots\uparrow\uparrow(\uparrow\uparrow)_{\ell'}\cdots\bullet\cdots}\,\\
\frac{1}{\pi}\arccos\big(-\tfrac{\tilde x(\ell')}{4\mathcal{J}t}\big)\,,&\text{for }\begin{cases}
\ket{\cdots \bullet\cdots(\uparrow\uparrow)_{\ell'}\uparrow\downarrow\cdots}\vee\ket{\cdots\uparrow\uparrow(\uparrow\downarrow)_{\ell'}\cdots\bullet\cdots}\\
\ket{\cdots \bullet\cdots(\uparrow\uparrow)_{\ell'}\downarrow\uparrow\cdots}\vee\ket{\cdots\uparrow\uparrow(\downarrow\uparrow)_{\ell'}\cdots\bullet\cdots}
\end{cases}\\
\frac{1}{\pi}\arccos\big(\tfrac{\tilde x(\ell')}{4\mathcal{J}t}\big)\,,&\text{for }\begin{cases}\ket{\cdots \bullet\cdots(\uparrow\downarrow)_{\ell'}\uparrow\uparrow\cdots}\vee\ket{\cdots \uparrow\downarrow(\uparrow\uparrow)_{\ell'}\cdots\bullet\cdots}\\
\ket{\cdots \bullet\cdots(\downarrow\uparrow)_{\ell'}\uparrow\uparrow\cdots}\vee\ket{\cdots\downarrow\uparrow(\uparrow\uparrow)_{\ell'}\cdots\bullet\cdots}
\end{cases}\\
0\,,&\text{for }\begin{cases}\ket{\cdots \bullet\cdots(\uparrow\downarrow)_{\ell'}\uparrow\downarrow\cdots}\vee
\ket{\cdots\uparrow\downarrow(\uparrow\downarrow)_{\ell'}\cdots\bullet\cdots}\\
\ket{\cdots \bullet\cdots(\downarrow\uparrow)_{\ell'}\uparrow\downarrow\cdots}\vee\ket{\cdots\downarrow\uparrow(\uparrow\downarrow)_{\ell'}\cdots\bullet\cdots}\\
\ket{\cdots \bullet\cdots(\downarrow\uparrow)_{\ell'}\downarrow\uparrow\cdots}\vee\ket{\cdots\downarrow\uparrow(\downarrow\uparrow)_{\ell'}\cdots\bullet\cdots}
\end{cases}
\end{cases} 
\ee
for the magnetisation, and
\be
\Bigl\langle\frac{\sigma_{2\ell'}^z-\sigma_{2\ell'-1}^z}{2}\Bigr\rangle_t\sim\begin{cases}
-\frac{1}{\pi}\arccos\big(\tfrac{\tilde x(\ell')}{4\mathcal{J}t}\big)\,,&\text{for }\ket{\cdots \bullet\cdots(\uparrow\uparrow)_{\ell'}\uparrow\downarrow\cdots}\vee
\ket{\cdots\uparrow\uparrow(\uparrow\downarrow)_{\ell'}\cdots\bullet\cdots}\\
-\frac{1}{\pi}\arccos(-\tfrac{\tilde x(\ell')}{4\mathcal{J}t})\,,&\text{for }\ket{\cdots \bullet\cdots(\uparrow\downarrow)_{\ell'}\uparrow\uparrow\cdots}\vee
\ket{\cdots\uparrow\downarrow(\uparrow\uparrow)_{\ell'}\cdots\bullet\cdots}\\
\frac{1}{\pi}\arccos(\tfrac{\tilde x(\ell')}{4\mathcal{J}t})\,,&\text{for }\ket{\cdots \bullet\cdots(\uparrow\uparrow)_{\ell'}\downarrow\uparrow\cdots}\vee
\ket{\cdots \uparrow\uparrow(\downarrow\uparrow)_{\ell'}\cdots\bullet\cdots}\\
\frac{1}{\pi}\arccos(-\tfrac{\tilde x(\ell')}{4\mathcal{J}t})\,,&\text{for }\ket{\cdots \bullet\cdots(\downarrow\uparrow)_{\ell'}\uparrow\uparrow\cdots}\vee
\ket{\cdots\downarrow\uparrow(\uparrow\uparrow)_{\ell'}\cdots\bullet\cdots}\\
\frac{2}{\pi}\arcsin(\tfrac{\tilde{x}(\ell')}{\mathcal{J} t})\,,&\text{for }\ket{\cdots \bullet\cdots(\downarrow\uparrow)_{\ell'}\uparrow\downarrow\cdots}\vee
\ket{\cdots\downarrow\uparrow(\uparrow\downarrow)_{\ell'}\cdots\bullet\cdots}\\
-1\,,&\text{for }\ket{\cdots \bullet\cdots(\uparrow\downarrow)_{\ell'}\uparrow\downarrow\cdots}\vee
\ket{\cdots\uparrow\downarrow(\uparrow\downarrow)_{\ell'}\cdots\bullet\cdots}\\
0\,,&\text{for }\ket{\cdots \bullet\cdots(\uparrow\uparrow)_{\ell'}\uparrow\uparrow\cdots}\vee
\ket{\cdots\uparrow\uparrow(\uparrow\uparrow)_{\ell'}\cdots\bullet\cdots}\\
1\,,&\text{for }
\ket{\cdots\bullet\cdots(\downarrow\uparrow)_{\ell'}\downarrow\uparrow\cdots}\vee
\ket{\cdots\downarrow\uparrow(\downarrow\uparrow)_{\ell'}\cdots\bullet\cdots}\end{cases} 
\ee
for the staggered magnetisation: in both cases $\tilde x(\ell')\sim x(2\ell')\sim x(2\ell'-1)$ is the coarse-grained position in the chain.

The currents of the magnetisation and of the staggered magnetisation
are obtained from the time derivative of $\sigma_\ell^z$ in the Heisenberg picture
\begin{align}
    \frac{{\rm d}}{{\rm d}t}\sigma^z_\ell(t)\Bigr|_{t=0}=-\mathcal{J} D_{\ell,\ell+2}\frac{1-\sigma_{\ell+1}^z}{2}+\mathcal{J} D_{\ell-2,\ell}\frac{1-\sigma_{\ell-1}^z}{2}\, ,
\end{align}
where $D_{\ell,m}=\sigma_\ell^x\sigma_m^y-\sigma_\ell^y\sigma_m^x$. In a jammed state, each term on the right-hand side vanishes. We can therefore conclude that the two currents approach zero in any locally quasi-jammed state in which the jamming condition is satisfied up to $O(1/t)$ corrections. 

The reader could be surprised at this result as we showed earlier that the corresponding charges have a nontrivial profile in the ballistic scaling limit $x\sim \mathcal{J} t$: in these situations, using the continuity equation, one could infer that also the differences of currents should be functions of $x/t$
\be
\partial_t q(\tfrac{x}{t})+\partial_x j(x,t)=0\quad\Rightarrow\quad j(x,t)-j(0,t)=\int_0^{x/t}{\rm d} \zeta q'(\zeta)\zeta\,.
\ee
Here, $q$ is the expectation value of the charge density and $j$ that of the corresponding current. 
This argument can however be applied only if the charge density is a (piece-wise) continuous function of $x/t$. In our case we have an additional microscopic dependence on the lattice position $q=q(\ell,t)$: the charge expectation value depends on the pattern of spins around the site $\ell$ -- cf.~\eqref{eq:asymptotic_magnetisation}. As we move from site to site, $q$ jumps between four smooth asymptotic curves that can be recognised only on large (ballistic) scales. The continuity equation on the chain then reads
\be
\partial_t q(\ell,t)=j(\ell+1,t)-j(\ell,t)
\ee
and, due to the microscopic dependence on $\ell$, it now becomes possible to have a nontrivial ballistic profile for the charge notwithstanding the corresponding current approaching zero. The macroscopic reconfiguration of the spin profiles after a localised perturbation is thus not a typical transport phenomenon: as shown in Section~\ref{sec:fluctuations}, the positions of the spins up only experience ``collective'' fluctuations of magnitude $O(1)$. The phenomenon is somewhat reminiscent of the travelling surface waves in which the particles of water undergo only a short-scale oscillatory motion. The distinguishing and perhaps most surprising feature of the phenomenon shown here is that it is a result of an instantaneous perturbation.
%%%
\begin{figure}[h!]
  \hspace{-1.5em}\centering
  \includegraphics[width=0.95\textwidth]{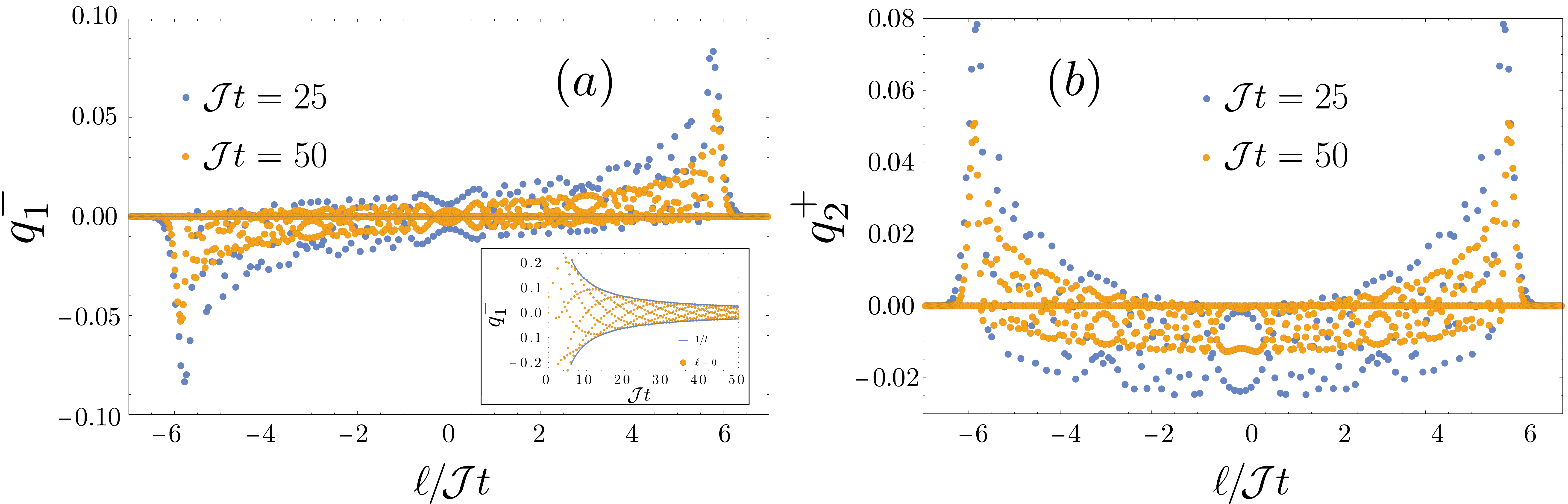}
  \caption{Spin current $q_1^-(\ell,t)$ and higher charge $q_2^+(\ell,t)$ for $\mathcal{J}t=25$ (blue points) and $\mathcal{J}t=50$ (orange points), obtained from the DMRG data for the initial state $\ket{\cdots\uparrow\uparrow\downarrow\uparrow\uparrow\downarrow\uparrow \downarrow \downarrow\uparrow\uparrow\downarrow\uparrow\uparrow\downarrow\cdots}$. In the large-time limit both profiles vanish as $1/t$, whereas the charges $q_1^+$ and $q_2^-$ are strictly zero during the time evolution.}
  \label{fig:Q1_minus_Q2_plus}
\end{figure}
%%%

Using the thermodynamic Bethe Ansatz description developed in Ref.~\cite{folded:2}, it can also be shown that the expectation values of the remaining (known) local conservation laws (reported in Refs~\cite{folded:1,folded:2}) asymptotically vanish. In the thermodynamic limit their expectation values per macrosite read~\cite{folded:2}
\begin{align}
\label{eq:charge_expect_value}
    q_n^\pm(x,t)=\int{\rm d}p\, \rho_{x,t}(p) \mathcal{Q}_n^\pm(p)\,,
\end{align}
where the root density $\rho_{x,t}(p)$ encodes the distribution of the occupied momenta at $x$ and $t$, while $\mathcal{Q}_n^+(p)=4\mathcal{J}\cos(n p)$ and $\mathcal{Q}_n^-(p)=4\mathcal{J}\sin(np)$ are the single-particle charge expectation values. The root density $\rho$ and the total density $\rho_{\rm t}$ of vacancies (i.e., particles and holes) in the momentum space satisfy $\int{\rm d}p\,\rho(p)=\xi^{-1}$ and $2\pi\rho_{\rm t}=1+\mu/\xi$, respectively,
where $\xi^{-1}$ is the particle density per macrosite, while $\mu$ is the thermodynamic limit of the ratio $\sum_{j}b_j(1-b_{j+1})/N$, $N$ being the number of spins up (see Ref.~\cite{folded:2}). In a jammed state of an infinite system one has $\xi+\mu=1$, whence $\int{\rm d}p\,\rho(p)=1/(1-\mu)$ and $\rho_{\rm t}=1/(2\pi[1-\mu])$~\cite{folded:1}. The occupancy ratio $n=\rho(p)/\rho_{\rm t}$ then satisfies $\int {\rm d}p\,n(p)=2\pi$ and, since $0\le n(p)\le 1$, one has $n(p)=1$. This implies that $\rho(p)$ is constant, whence all local charge expectation values, expect for the magnetisation (and the staggered one, albeit for a different reason\footnote{The staggered magnetisation is not related to the root density $\rho$, so the argument above does not pertain to it.}), vanish:
\begin{align}
    q_n^\pm(x,t)=0,\quad\text{for all $n>0$}\, .
\end{align}

A similar conclusion can be drawn for the currents, whose expectation values read
\begin{align}
    j_n^\pm(x,t)=\int{\rm d}p\, \rho_{x,t}(p)v_{x,t}(p) \mathcal{Q}_n^\pm(p) + \text{const.}\,,
\end{align}
where $v(p)=-4 \mathcal{J} \sin p -4\mathcal{J}\mu\int\frac{\mathrm d k}{2\pi}n(k)(\sin k-\sin p)$ is the dressed velocity~\cite{folded:2}. By restricting to jammed states, we immediately see that all currents become independent of the state and hence approach trivially constant profiles in locally quasi-jammed states like those emerging in our setting.

\subsection{Entanglement in a weakly interacting setting}
\label{s:special_protocol}
In this section we identify both classical and pure quantum contributions to the two-point spin correlation functions. We focus on a weakly interacting scenario in which the background consists of jammed particles that are of the same species almost everywhere except for those in a special region. In the latter, both species of particles are present in the maximal number.
%, in which the spin profiles, the jamming condition, the entanglement between a single spin and the rest of the chain, as well as the entanglement between any two spins can be computed explicitly for all $t>0$. 
In other words, the initial jammed state is obtained from the N\'eel state by flipping up a sequence of $M$ spins down. We denote the corresponding macrosites by $m',m'+1,\ldots,m'+M-1$, so we have%: the macrosite description is suitable due to the two-site shift invariance of the N\'eel state. The state now reads
\begin{equation}
\label{eq:weakly_interacting_state}
    \ket{\Psi(0^-)}=\ket{\cdots\downarrow\uparrow\downarrow\uparrow \underset{m'}{\underline{\uparrow\uparrow}}\uparrow\uparrow \cdots \uparrow\uparrow \underset{m'+M-1}{\underline{\uparrow\uparrow}} \downarrow\uparrow\downarrow\uparrow\cdots}\,,
\end{equation}
where we are going to assume $m'>0$. Note that for this specific state $M=\sum_{j}b_j(1-b_{j+1})$. In the light of the Bethe Ansatz solution of Ref.~\cite{folded:1}, a nonzero sum $\sum_{j}b_j(1-b_{j+1})$ is always associated with interaction. Since, however, the sum is finite and we are in the thermodynamic limit, the state~\eqref{eq:weakly_interacting_state} can be considered weakly interacting. By flipping the spin in position $0$ according to the protocol described in Section~\ref{s:protocol}, we end up with the state
\begin{equation}
\label{eq:weakly_interacting_state_2}
\ket{\Psi(0)}=    \ket{\cdots\downarrow\uparrow\underset{0}{\underline{\downarrow\downarrow}}\downarrow\uparrow \cdots\downarrow\uparrow\downarrow\uparrow 
    \underset{m'}{\underline{\uparrow\uparrow}}\uparrow\uparrow \cdots \uparrow\uparrow 
    \underset{m'+M-1}{\underline{\uparrow\uparrow}} \downarrow\uparrow\downarrow\uparrow\cdots}\,,
\end{equation}
which is no longer jammed, but evolves in time towards a locally quasi-jammed state. One can indeed readily show (see Appendix~\ref{app:special_protocol})
\begin{equation}
    P_{\downarrow\downarrow}(\ell) =
    \left\{\begin{aligned}
    &J^2_{\lceil\ell/2\rceil}(4 \mathcal{J} t)\,, &&\text{if } \lfloor\ell/2\rfloor<m'-1 \\
    &J^2_{\ell-m'+2}(4 \mathcal{J} t)\,, &&\text{if } m'-1\le \lfloor\ell/2\rfloor <m'+M-2 \\
    &J^2_{\lceil\ell/2\rceil+M}(4 \mathcal{J} t)\,, &&\text{if } \lfloor\ell/2\rfloor\ge m'+M-2\,, \\
    \end{aligned}\right.
\end{equation}
hence the jamming condition is asymptotically satisfied up to $O(1/t)$ corrections -- cf.~\eqref{eq:LQJcond}. 

\begin{figure}[t!]
  \hspace{-1.5em}\centering
  \includegraphics[width=0.65\textwidth]{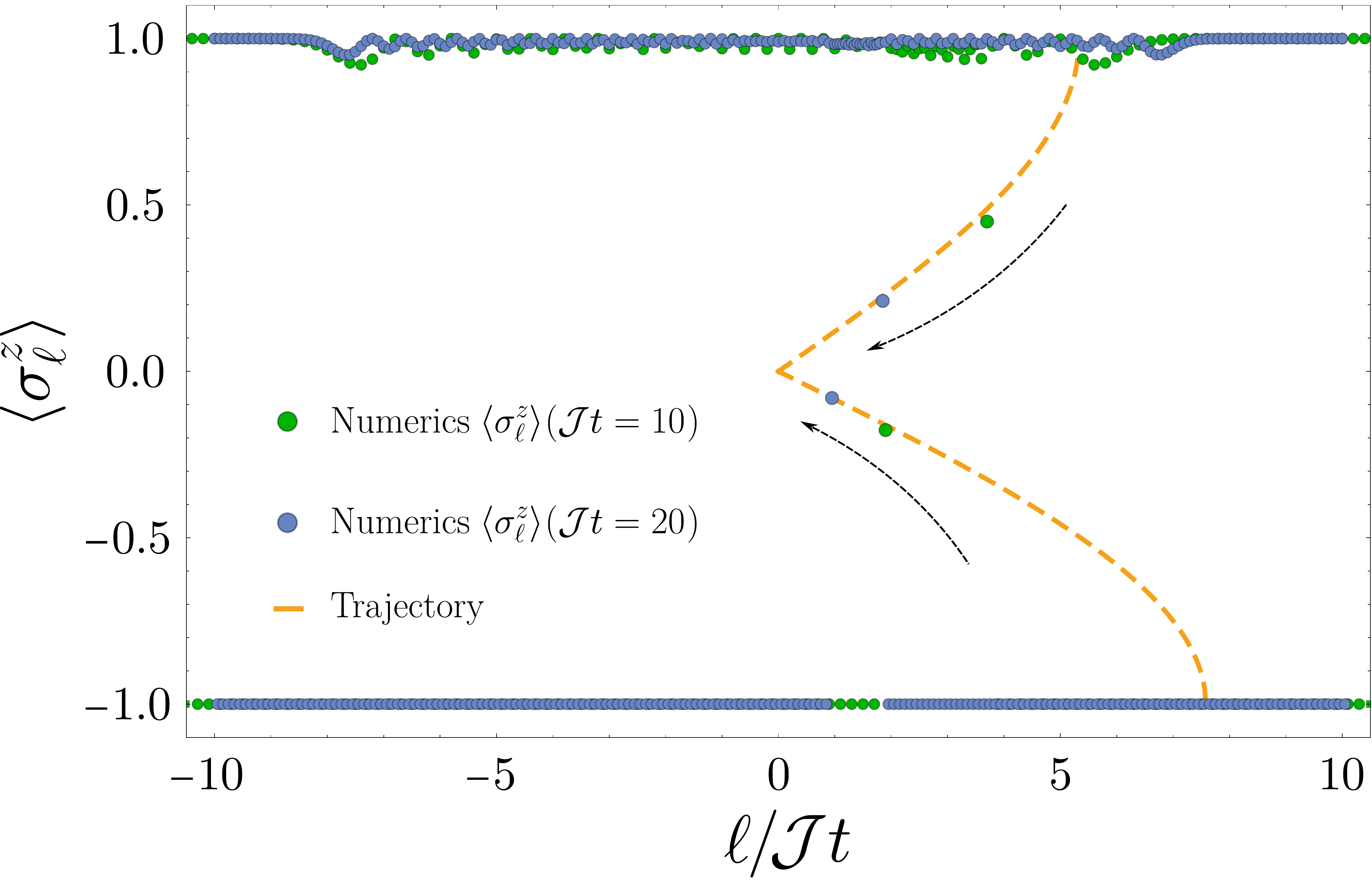}
  \caption{Magnetisation profile $\braket{\sigma^z_{\ell}}_t$ for the initial state described in Eq.~\eqref{eq:weakly_interacting_state_2} with $m'=9$ and $M=10$. Colors of the points correspond to different fixed times $\mathcal{J}t=10$ (green) and $\mathcal{J}t=20$ (blue). The spins at sites $2m'-3$ and $2m'+2M-3$ are the only ones for which the $z$-components of the magnetisation are not asymptotically equal to $\pm 1$. Instead, as time $t$ is increased, they move towards the zero ray (time increases while the lattice positions are fixed) along the trajectories (orange dashed lines) given in Eq.~\eqref{eq:special_trajectories}. %The trajectory of the outer spin depends on the ratio $M/m'$, which is fixed by the initial configuration of spins. 
  }
  \label{fig:Neel_magnetization}
\end{figure}

For the local magnetisation we obtain
\begin{align}
\begin{aligned}
    \label{eq:Neel_magnetization}
    \braket{\sigma^z_{2\ell'-1}}_t&=
    \left\{
    \begin{aligned}
    &2f(m'+2M-2,t)-1 \,,&&\text{if } \ell'=m'+M-1
    \\
    &1-2f(m',t) \,,&&\text{if }  \ell'=m'-1
    \\
    &1-2J^2_{m'-2\ell'-1}(4 \mathcal{J} t)-2J^2_{m'-2\ell'}(4 \mathcal{J}t) \,,
    \phantom{ii}
    && \text{if } m'\le\ell'\le m'+M-2
    \\
    &-1 \,,&& \text{otherwise}\,,
    \end{aligned}
    \right. \\
    \braket{\sigma^z_{2\ell'}}_t&=
    \left\{
    \begin{aligned}
    &1 
    -2J^2_{\ell'}(4 \mathcal{J} t) 
    \,,&&\text{if } \ell'<m'-1
    \\&1 
    -2J^2_{\ell'+M}(4 \mathcal{J} t)
    \,,&&\text{if } \ell'>m'+\Delta-2
    \\&1 
    -2 J^2_{m'-2\ell'-2}(4 \mathcal{J} t)
    -2 J^2_{m'-2\ell'-1}(4 \mathcal{J} t)
    \,,&& \text{if } m'-1\le \ell'\le m'+M-2
    \,,
    \end{aligned}
    \right.
\end{aligned}
\end{align}
where $f(x,t)$ is defined in Eq.~\eqref{eq:f_function}, while $\braket{\sigma^x_\ell}_t=0$ and $\braket{\sigma^y_\ell}_t=0$ (see Appendix~\ref{app:special_protocol_SpinProfiles} for details).  Note that, except for the sites $2m'-3$ and $2m'+2M-3$, the long-time limit of the expectation value of $\sigma^z$ is either $1$ or $-1$. 
Using the asymptotic expression for the function $f$, reported in Eq.~\eqref{eq:f_function}, we instead obtain
\begin{equation}
\label{eq:special_trajectories}
\begin{aligned}
    \braket{\sigma^z_{2m'-3}}_t&\sim -\frac{2}{\pi}\arcsin\Bigl(\frac{m'}{4 \mathcal{J} t}\Bigr),\qquad
    \braket{\sigma^z_{2m'+2M-3}}_t&\sim \frac{2}{\pi}\arcsin\Bigl(\frac{m'}{4 \mathcal{J} t}\Big[1+2\frac{M-1}{m'}\Big]\Bigr)\,,
\end{aligned}
\end{equation}
for the $z$-components of the spins at the sites $2m'-3$ and $2m'+2M-3$ (see Fig.~\ref{fig:Neel_magnetization}). This implies that, in the infinite-time limit, the reduced density matrix of any spin with site index $j\nin\{2m'-3,2m'+2M-3\}$ is pure. The reduced density matrix of each of the two spins at sites $2m'-3$ and $2m'+2M-3$ is instead maximally mixed (proportional to the identity). Hence, the spins $2m'-3$ and $2m'+2M-3$ are maximally entangled with the rest of the system (the entanglement entropy of their reduced density matrices is maximal).

% The entanglement is computed from the reduced density matrix of the subsystem in consideration, constructed from
% \begin{equation}
%     \rho_{i_1,\ldots,i_n}(t) = 2^{-n}
%     \sum_{\alpha_1,\ldots,\alpha_n\in\{0,x,y,z\}}
%     \braket{\sigma^{\alpha_1}_{i_1}\ldots\sigma^{\alpha_n}_{i_n}}_t
%     \sigma^{\alpha_1}_{i_1}\ldots\sigma^{\alpha_n}_{i_n}\,,
% \end{equation}
% where the subsystem is formed by the spins labeled by indices $i_1,\ldots,i_n$.

On the other hand, the large-time limit of the correlation $\braket{\sigma^z_{2m'-3}\sigma^z_{2m'+2M-3}}_t$ is $-1$, i.e., the two spins are asymptotically maximally correlated. One might be tempted to conclude that these two spins are entangled with one another even in the infinite-time limit, i.e., that they are in a cat-state. Instead, it can be shown that their entanglement goes to zero as $\log(\mathcal{J}t)/(\mathcal{J}t)^2$, as does the entanglement between any other two spins. To see this, we compute the entanglement of formation~\cite{Wootters1997,Wootters1998} $E$ between two given spins:
\begin{equation}
    E(\hat \rho)=h\Bigl(\frac{1+\sqrt{1-C^2(\hat \rho)}}{2}\Bigr)\,.
\end{equation}
Here,  $\hat \rho$ is the reduced density matrix of the two spins,
\begin{equation}
    h(x)= -x\log_2x-(1-x)\log_2(1-x)\, ,
\end{equation}
while $C(\hat \rho)$ is the so-called concurrence, defined as
\begin{equation}
    C(\hat \rho)=\max\{0,\lambda_1-\lambda_2-\lambda_3-\lambda_4\},
\end{equation}
where $\lambda_j$ are the square roots of the eigenvalues of $\hat \rho(\sigma^y\otimes\sigma^y)\hat \rho^\ast(\sigma^y\otimes\sigma^y)$ in descending order, and $\hat\rho^\ast$ is the complex conjugate of $\hat\rho$. Even though $\hat\rho(\sigma^y\otimes\sigma^y)\hat\rho^\ast(\sigma^y\otimes\sigma^y)$ is not necessarily self-adjoint, its eigenvalues are real and non-negative (it is the product of two positive semi-definite matrices).
%%%
%%%
Note that the concurrence (and hence the entanglement) is automatically zero whenever the density matrix is diagonal because the matrix $\hat\rho(\sigma^y\otimes\sigma^y)\hat{\rho}^\ast(\sigma^y\otimes\sigma^y)$ has two pairs of degenerate positive eigenvalues. One can show that the off-diagonal elements of the two-spin reduced density matrix $\hat \rho$ go to zero as $1/t$ for large times, thus lifting  the degeneracy of the matrix $\hat \rho(\sigma^y\otimes\sigma^y)\hat{\rho}^\ast(\sigma^y\otimes\sigma^y)$ by a term of order $O(1/t)$. This makes the entanglement of formation approach zero as $\log(\mathcal{J}t)/(\mathcal{J}t)^2$.
Nonetheless, the two spins at sites $2m'-3$ and $2m'+2M-3$ are special, since they are maximally correlated: in the infinite-time limit their reduced density matrix reads
\begin{equation}
    \hat{\rho}\xrightarrow{t\rightarrow\infty} \frac{\ket{\uparrow\downarrow}\bra{\uparrow\downarrow} +\ket{\downarrow\uparrow}\bra{\downarrow\uparrow}}{2}\,,
\end{equation}
which is a classical mixture. Note that the macro-distance $M$ between the correlated spins can be arbitrarily large.

\begin{figure}[t!]
    \centering
%     \begin{subfigure}[b]{0.3\textwidth}
%     \includegraphics[width=\textwidth]{img/DeltaProtDelta7size120emme18time1point5.png}
%     \caption{$t=1.5$}
%   \end{subfigure}
%     \begin{subfigure}[b]{0.3\textwidth}
%     \includegraphics[width=\textwidth]{img/DeltaProtDelta7size120emme18time5point1.png}
%     \caption{$t=5.1$}
%   \end{subfigure}
%     \begin{subfigure}[b]{0.3\textwidth}
%     \includegraphics[width=\textwidth]{img/DeltaProtDelta7size120emme18time9point9.png}
%     \caption{$t=9.9$}
%   \end{subfigure}
%   \begin{subfigure}[b]{0.05\textwidth}
%   \begin{tikzpicture}
% \node (scale) at (0,0) {\includegraphics[height=4.6\textwidth]{./img/scale.pdf}};
% \node[above =1cm] at (.4,0) {$0.6$};
% \node[below =1.4cm] at (.4,0) {$0.0$};
% \node at (.4,-.2) {$0.3$};
% \end{tikzpicture}
%  \vspace{.01\textwidth}
  
%   \caption*{ }
%   \end{subfigure}
  
    \begin{tikzpicture}%[scale=\textwidth/12.8cm]
\node at (-12.6,0) {\includegraphics[width=.3\textwidth]{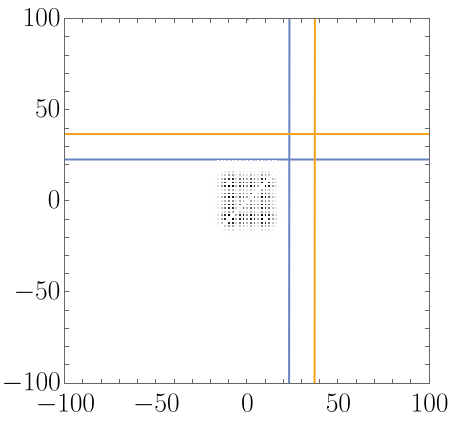}};
\node at (-12.6,-2.7) {(a) $\mathcal{J}t=1.5$};
\node at (-7.8,0) {\includegraphics[width=.3\textwidth]{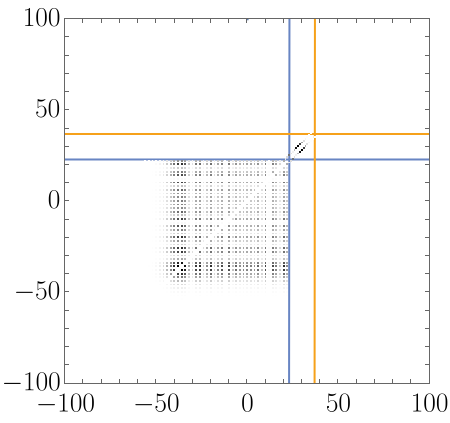}};
\node at (-7.8,-2.7) {(b) $\mathcal{J}t=5.1$};
\node at (-3,0) {\includegraphics[width=.3\textwidth]{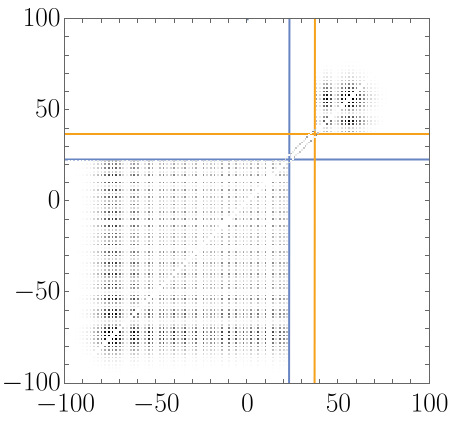}};
\node at (-3,-2.7) {(c) $\mathcal{J}t=9.9$};
\node (scale) at (-.4,0) {\includegraphics[height=.2\textwidth]{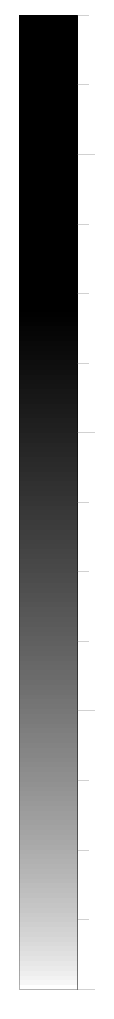}};
\node[above =.84cm] at (.1,0) {$0.6$};
\node[below =1.2cm] at (.1,0) {$0.0$};
\node at (.1,-.18) {$0.3$};
\end{tikzpicture}

    \caption{Spin-spin entanglement for an initial state with $m'=12$, $M=9$ at different times, rescaled by $\log (\mathcal{J}t)/(\mathcal{J}t)^2$. 
    The numbers on both axes refer to spin indices: the element $(i,j)$ of the plot characterises the entanglement between the spins at the sites $i$ and $j$. The plot is symmetric w.r.t. the bisector, i.e., the line containing the points $(i,i)$, since the entanglement between the spins $i$ and $j$ equals that between the spins $j$ and $i$. 
    The blue (orange) lines refer to the spin $2m'-1$ (resp. $2m'+2M-5$), i.e., the one close to the start (resp. end) of the domain wall of spins up. They delimit the nonzero entanglement regions. Panels (a), (b), and (c) show the spread of the entanglement before the domain of spins up is reached by the light cone, during the spread of the entanglement through the domain, and after the edge of the light cone has passed through the domain, respectively.
    }
    \label{fig:delta_protocol_different_times}
\end{figure}
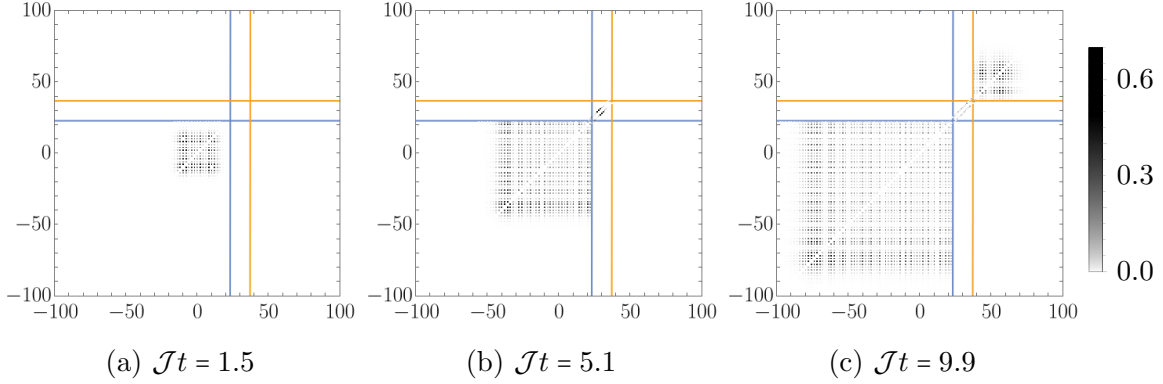

In general, the spread of the rescaled spin-spin entanglement with time is shown in Fig.~\ref{fig:delta_protocol_different_times}. At $t=0$ there are no entangled spins (the initial state is a product state). For $t>0$ the region containing entangled spins grows linearly in time, consistently with the light-cone dynamics. At first the entanglement spreads from the initial position of the impurity in all directions -- cf. Fig.~\ref{fig:delta_protocol_different_times}(a) -- with the maximal velocity $8J$. %Before the edge of the light cone reaches the macrosite $m$ the entanglement spreads in all directions -- cf. Fig.~\ref{fig:delta_protocol_different_times}(a). 
When the edge of the light-cone reaches macro-site $m$, a ``leakage'' of the entanglement between the wedges occurs through a corridor, in which only the next-nearest neighbour spins become entangled -- cf. Fig.~\ref{fig:delta_protocol_different_times}(b,c).
We observe that the couples of spins $(i,j)$ with $i,j<2m'-1$ are not affected by the presence of the domain wall in the initial state: their behaviour is the same as if the initial state were a N\'eel state. Instead, the entanglement properties of the pairs of spins $(i,j)$ with $i,j>2m'+M-5$ are shifted: their entanglement corresponds to that in the N\'eel state in which the impurity is initially located at macro-site $-M$. Finally, in our case, all the spins outside those two regions are decoupled unless they are next-nearest neighbours, which is not the case when the dynamics start from the N\'eel state.

\section{From the dual folded XXZ to the XXZ  model}

\begin{figure}[t!]
  %\hspace{-.5em}
  \centering
  \includegraphics[width=0.95\textwidth]{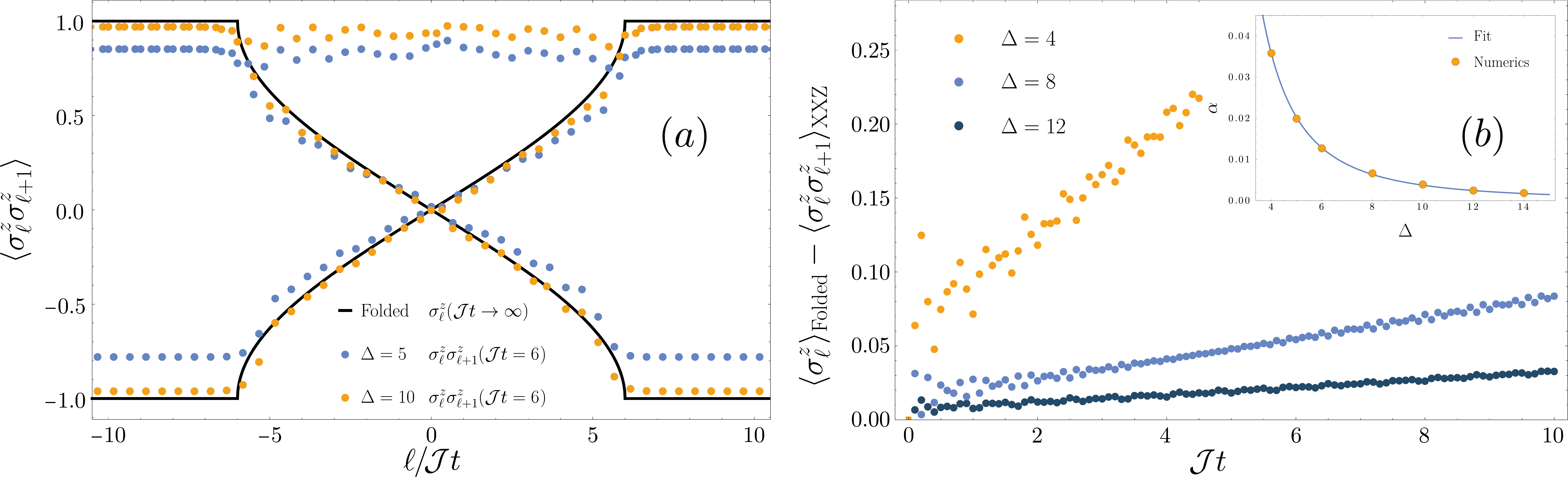}
  \caption{Panel (a) shows correlation function $\braket{\sigma^z_\ell \sigma^z_{\ell+1}}_t$ in the Heisenberg XXZ model (blue and orange points) compared to the magnetization of the folded model Eq.~\eqref{eq:asymptotic_magnetisation} (black line). Data points quickly converge to the prediction of the folded model for a relatively large values of $\Delta$, whereas for small anisotropy they exhibit a disagreement with the prediction during the whole time evolution. Panel (b) shows the difference $\braket{\sigma^z_\ell}_{\rm folded} -\braket{\sigma^z_\ell \sigma^z_{\ell+1}}_{\rm XXZ}$ outside of the light cone as a function of time. For short times the predictions in both models linearly deviate from each other. In the inset we plot a slope $\alpha$ of this difference as a function of the anisotropy $\Delta$. Data points are fitted with $a\Delta^{-2} + b\Delta^{-3}$ , where $a\approx0.241$, $b\approx1.31$.}
  \label{fig:XXZ_vs_Folded}
\end{figure}

The dual folded XXZ model was originally envisioned as an effective model describing the large-anisotropy limit of the Heisenberg XXZ spin-$1/2$ chain. In this respect it is natural to wonder how well it acts as such. Specifically, is there a sign of the measurement catastrophe studied herein and in Ref.~\cite{measurement2021} in the XXZ model with large anisotropy? To answer this question, one needs to consider the inverse of the duality transformation and invert an additional unitary transformation that acts on the initial state and characterises the so-called ``folded picture'' (see Ref.~\cite{folded:1}). The latter unitary transformation only contributes $O(1/\Delta)$ corrections and will be neglected in the following discussion. The inverse duality transformation maps $\sigma_\ell^z\mapsto \sigma_\ell^z\sigma_{\ell+1}^z$, so it essentially acts as follows:
\begin{align}
\label{eq:spin_mapping}
    \ket{\cdots\uparrow\uparrow\downarrow\uparrow\uparrow\downarrow\uparrow\downarrow\downarrow\uparrow\uparrow\downarrow\uparrow\uparrow\downarrow\cdots}\xrightarrow[\text{transformation}]{\text{inverse duality}} \ket{\cdots\uparrow\uparrow\uparrow\downarrow\downarrow\downarrow\uparrow\uparrow\downarrow\uparrow\uparrow\uparrow\downarrow\downarrow\downarrow\uparrow\cdots}\, .
\end{align}
Time-evolving this initial state with the Heisenberg Hamiltonian $H_{\rm XXZ}(\Delta)$ is expected to yield a scaling profile of $\braket{\sigma_\ell^z\sigma_{\ell+1}^z}_t$ converging, as $\Delta$ is increased, to the one of $\braket{\sigma_\ell^z}_t$ in the dual folded XXZ model. This is indeed observed in numerical simulations based on tensor network techniques -- cf. Fig.~\ref{fig:XXZ_vs_Folded}, and supports the validity of the strong-coupling expansion described in Ref.~\cite{folded:1}.
Besides that, transforming back to XXZ helps us see why the local measurement could have such a relevant effect: in the XXZ setting, it creates a domain-wall structure. If the two halves of the chain had different integrals of motion, the theory of generalised hydrodynamics~\cite{castro16,bertini16,vir17} would predict nontrivial ballistic profiles. This is not the case. One could then wonder whether we are seeing the effects of quasi-conserved operators that are exactly conserved only in the large-anisotropy limit. So far, our investigations have been inconclusive, so the question is still open.

%%%
%%%

\section{Discussion}
The simplicity of the calculations carried out in this paper is rather surprising in view of the fact that we are dealing with a genuinely interacting model. It goes even beyond the simplicity of the Bethe equations of the dual folded XXZ model~\cite{folded:1}. It seems rather related to the setting considered -- that of locally perturbed jammed states, in which particles are closely packed. Such a scenario can be mapped into the problem of a free particle moving in a jammed background. We have exploited this mapping to solve the measurement catastrophe discussed in Ref.~\cite{measurement2021}, i.e., the macroscopic change of spin profiles after a localised perturbation. We have established that no genuine spin transport occurs: the particles' positions exhibit small fluctuations around their expectation values, and the current of the local magnetisation is asymptotically zero. This is a result of the emergent scale separation in which the smooth asymptotic scaling of the sublattice local magnetisation coexists with an underlying discontinuous dependence of $\braket{\sigma^z_\ell}$ on the position $\ell$. Interactions play a key role in the setting we studied: the phenomenon is not observed if particles do not interact. On the other hand, a finite number of interacting particles is enough to trigger the effect. In particular, perturbing a weakly interacting initial state enables isolation of classically correlated pairs of spins. 

Jammed states typically arise in kinetically constrained models, and it remains unclear whether the measurement catastrophe described in this work is a universal feature of such models~\cite{garrahan2010kinetically}. 

During the finalisation of this work and in the attempt to unveil the physical mechanism behind the measurement catastrophe, one of us has realised that an analogous sensitivity to localised perturbations can be present also after global quenches~\cite{Fagotti2021Global}. In that case the existence of infinitely many ``semilocal charges'' was identified as crucial. That mechanism, however, does not seem to account for the phenomenon discussed in this paper, which appears to be more related to the exceptionality  of the jammed sector,  consisting of states with memory. Indeed, to the best of our knowledge the dual folded Hamiltonian only has a single semilocal charge, first identified in Ref.~\cite{folded:1}, which corresponds to the total spin along the $z$-axis when mapping back to the XXZ Heisenberg model. That charge has a nontrivial profile in some of the examples that we considered, but it is irrelevant in others, including in the example investigated in Ref.~\cite{measurement2021}.

%%%%%%%%%%%%%%%%
\section*{Acknowledgements}%
%%%%%%%%%%%%%%%%

%%%%%%%%%%%%%%%%%%
\paragraph{Funding information.} %
%%%%%%%%%%%%%%%%%%
This work was supported by the European Research Council under the Starting Grant No. 805252 LoCoMacro.

%\begin{figure}[h!]
% \centering
%  \begin{subfigure}[b]{0.48\textwidth}
%    \includegraphics[width=\textwidth]{XXZ5.pdf}
%    \caption{$\Delta = 5$. Deviations become visible at relatively short times $Jt \approx \Delta = 5$}
%  \end{subfigure}
%  \begin{subfigure}[b]{0.48\textwidth}
%    \includegraphics[width=\textwidth]{XXZ10.pdf}
%    \caption{$\Delta = 10$. Numerical data collapses with analytical scaling curve}
%  \end{subfigure}
%  \caption{$\langle \sigma^z_\ell \sigma^z_{\ell+1}\rangle_{\rm xxz} (\ell/t)$ profile in comparison with $\langle \sigma^z_r \rangle$ of the Folded model}
%  \label{fig:XXZ}
%\end{figure}

%\begin{figure}[h!]
%\centering
%\includegraphics[width=0.8\textwidth]{XXZ4_8_16.pdf}
%\caption{$\langle \sigma^z_\ell \sigma^z_{\ell+1}\rangle_{\rm xxz} (\ell/t)$ profile for different values of $\Delta = 4, 8, 16$}
%\label{fig:XXZ_Large_Delta}
%\end{figure}

\bibliography{references.bib}
\nolinenumbers

\clearpage

\begin{appendix}
\renewcommand\thefigure{\thesection.\arabic{figure}}
\setcounter{figure}{0}

\numberwithin{equation}{section}

\section{Proofs}\label{s:proof}
These sections contain the proofs of the equations stated in the main part of the text. 

\subsection{Time evolution of the state and \texorpdfstring{$\braket{\sigma_\ell^z}_t$}{TEXT}}
\label{app:time_evolution_of_state}

Recall that the state $\ket{n;\underline b^{(0)}}$ is defined in such a way that the impurity lies on the immediate right-hand side of the particle (spin up) with index $n$. The state is jammed everywhere, except at the position of the impurity, which consists of either two or three consecutive spins down. The width of the impurity, i.e., the number of spins down that it contains, is not preserved under the action of the Hamiltonian. Nevertheless, the types of particles represented by spins up are well defined and their sequence is preserved, as shown in Ref.~\cite{folded:1} (e.g., the particle corresponding to a spin up at the site $\ell$ is of the type $b=2\lceil\ell/2\rceil-\ell$).

Let us define the impurity creation operator $c_n^\dagger$ and the reference state $\ket{\emptyset;\underline{b}^{(0)}}$, so that
\begin{align}
    c_n^\dagger\ket{\emptyset;\underline{b}^{(0)}}:=\ket{n;\underline b^{(0)}}\, .
\end{align} 
Since the background $\underline{b}^{0}$ is preserved under the action of the Hamiltonian~\eqref{eq:Hamiltonian}, the impurity effectively behaves as a freely hopping fermion. States $\ket{n;\underline b^{(0)}}$ span an invariant subspace, on which the dynamics can be readily solved. Employing a diagonalisation procedure for a free hopping Hamiltonian in the thermodynamic limit, we obtain
\begin{align}
\begin{gathered}
    e^{-i H t}c^\dagger_{\ell}\ket{\emptyset;\underline{b}^{(0)}}=\sum_n \int\frac{{\rm d} k}{2\pi}e^{-i k (n-\ell)-i\varepsilon(k)t}c_n^\dagger\ket{\emptyset;\underline{b}^{(0)}}=\sum_n (-i)^{n-\ell} J_{n-\ell}(4\mathcal{J} t)\ket{n;\underline{b}^{(0)}}
\end{gathered}
\end{align}
for the evolution of the state $\ket{\ell;\underline{b}^{(0)}}\equiv c_\ell^\dagger\ket{\emptyset;\underline{b}^{(0)}}$. Here, the energy dispersion relation reads $\varepsilon(k)=4\mathcal{J}\cos k$, and $J_n(x)$ are Bessel functions. In particular, the evolution of the initial state $\ket{\Psi(0)}=\ket{0;\underline{b}^{(0)}}\equiv c_0^\dagger\ket{\emptyset;\underline{b}^{(0)}}$ in the protocol described in the main text reads
\begin{align}
    \ket{\Psi(t)}=e^{-i H t}c^\dagger_{0}\ket{\emptyset;\underline{b}^{(0)}}=\sum_n (-i)^{n} J_{n}(4\mathcal{J} t)\ket{n;\underline{b}^{(0)}}\,.
\end{align}

Consider now the expectation value $\braket{\sigma_\ell^z}_t=\bra{\Psi(t)}\sigma_\ell^z\ket{\Psi(t)}$, where $\ell$ is the site on the lattice. We have
\begin{align}
\label{eq:average_app}
    \braket{\sigma_\ell^z}_t=\sum_{n,m}\int\frac{{\rm d} k}{2\pi}\int\frac{{\rm d} p}{2\pi}e^{ikn-ipm}e^{i[\varepsilon(k)-\varepsilon(p)]t}\bra{n;\underline{b}^{(0)}}\sigma_\ell^z\ket{m;\underline{b}^{(0)}}.
\end{align}
The operator $\sigma_\ell^z$ is diagonal in the basis $\ket{n;\underline b^{(0)}}$, therefore the only non-vanishing matrix elements are $\braket{n;\underline b^{(0)}|\sigma_\ell^z|n;\underline b^{(0)}}$, whence
\begin{align}
\label{eq:bessel_formula}
    \braket{\sigma_\ell^z}_t=\sum_{n}|J_{n}(4 \mathcal{J} t)|^2\bra{n;\underline{b}^{(0)}}\sigma_\ell^z\ket{n;\underline{b}^{(0)}}\,.
\end{align}
In Eq.~\eqref{eq:bessel_formula} $\sigma_\ell^z$ can be substituted by any operator $D$, diagonal in the basis of states $\ket{n;\underline{b}^{(0)}}$.
For observables $A$ that are not diagonal in this basis, we instead have
\begin{align}
\label{eq:off_diagonal}
    \braket{A}_t=\sum_{n_1,n_2} e^{i\frac{\pi}{2}(n_1-n_2)}
    J_{n_1}(4 \mathcal{J}t) \overline{J_{n_2}(4 \mathcal{J} t)}
    \bra{n_1;\underline{b}^{(0)}}A\ket{n_2;\underline{b}^{(0)}}\,.
\end{align}

\subsection{Equal and non-equal time correlations}
\label{app:correlators}

Let $\langle\bullet\rangle\equiv \langle\bullet\rangle_{t=0}$ denote the average in the initial state $\ket{\Psi(0)}=\ket{0;\underline{b}^{(0)}}$. In this section we report different-time correlation functions of operators that are diagonal in the basis of states $\ket{n;\underline{b}^{(0)}}$. In particular, the general two-point correlation function of two diagonal operators $D_1$ and $D_2$ reads
\begin{align}
\begin{gathered}
    \braket{D_1(t_1)D_2(t_2)}=\sum_{m,n}J_{m}(4\mathcal{J}t_1)J_{m-n}(4\mathcal{J} [t_1-t_2])J_{n}(4\mathcal{J}t_2)\times\\
    \times\braket{m;\underline{b}^{(0)}|D_1|m;\underline{b}^{(0)}}\braket{n;\underline{b}^{(0)}|D_2|n;\underline{b}^{(0)}}\,.
\end{gathered}
\end{align}
At equal times the correlation function becomes
\begin{align}
\begin{gathered}
    \braket{D_1(t)D_2(t)}=\sum_{n}|J_{n}(4\mathcal{J} t)|^2
    \braket{n;\underline{b}^{(0)}|D_1|n;\underline{b}^{(0)}}\braket{n;\underline{b}^{(0)}|D_2|n;\underline{b}^{(0)}}\, ,
\end{gathered}
\end{align}
while at $t_2=0$ it factorises
\begin{align}
    \braket{D_1(t)D_2}=\big(\sum_m|J_{m}(4\mathcal{J}t)|^2\bra{m;\underline{b}^{(0)}}D_1\ket{m;\underline{b}^{(0)}}\Big)\bra{0;\underline{b}^{(0)}}D_2\ket{0;\underline{b}^{(0)}}=\braket{D_1}_t\braket{D_2}\, .
\end{align}

\subsection{Matrix elements of \texorpdfstring{$\sigma_\ell^z$}{TEXT}}
\label{app:matrix_elements}

In this section we compute the diagonal matrix elements $\braket{n;\underline b^{(0)}|\sigma_\ell^z|n;\underline b^{(0)}}$.

\subsubsection{Increasing \texorpdfstring{$n$}{TEXT}: the right-moving impurity}

First we consider the case of increasing $n$, so that the impurity passes from the left- to the right-hand side of the site $\ell$. The local Hamiltonian density 
\begin{align}
    h_{\ell,\ell+1,\ell+2}=\mathcal{J}\frac{1-\sigma^z_{\ell+1}}{2}(\sigma_\ell^x\sigma_{\ell+2}^x+\sigma_\ell^y\sigma_{+2}^y) 
\end{align}
acts on three subsequent sites. The impurity is identified with two or three consecutive spins down: once it reaches the left-hand side of the site $\ell$, two spins down occupy the sites $\ell-2$ and $\ell-1$. Three subsequent actions of the local Hamiltonian density are then required for it to pass to the right-hand side of the site $\ell$. Specifically, the actions are those of $h_{\ell-2,\ell-1,\ell}$, $h_{\ell-1,\ell,\ell+1}$, and $h_{\ell,\ell+1,\ell+2}$ (only two steps are needed, for example, when a spin down already occupies the $\ell$-th site). To determine the matrix element of $\sigma^z_\ell$ we thus need to consider eight processes, determined by the spins on sites $\ell$, $\ell+1$, and $\ell+2$ on the right-hand side of the right-moving impurity. Three of these processes do not occur, since the state is jammed everywhere but at the position of the impurity itself: we can not have $\uparrow\downarrow\downarrow$, $\downarrow\downarrow\uparrow$, or $\downarrow\downarrow\downarrow$ on sites $\ell$, $\ell+1$, and $\ell+2$, while the impurity is on the left-hand side of the site $\ell$. The remaining five processes that determine the matrix elements $\braket{n;\underline b^{(0)}|\sigma_\ell^z|n;\underline b^{(0)}}$ are presented in Fig.~\ref{fig:matrix_elements}, which serves as a proof of the following rules determining the matrix element:
\begin{align}
\label{eq:dynamical_rules_app}
\braket{n;\underline b^{(0)}|\sigma_{\ell>0}^z|n;\underline b^{(0)}}=\begin{cases}
1-2\delta_{n,j_R}-2\delta_{n,j_R+1}\,,&\text{if }\ket{0;\underline b^{(0)}}=\ket{\cdots\uparrow^{(0)}\bullet\uparrow^{(1)}\cdots\uparrow^{(j_R)}_\ell\uparrow\uparrow\cdots}\\
1-2\delta_{n,j_R}\,,&\text{if }\ket{0;\underline b^{(0)}}=\ket{\cdots\uparrow^{(0)}\bullet\uparrow^{(1)}\cdots\uparrow^{(j_R)}_\ell\downarrow\uparrow\cdots}\\
1-2\theta(n\geq j_R)\,,&\text{if }\ket{0;\underline b^{(0)}}=\ket{\cdots\uparrow^{(0)}\bullet\uparrow^{(1)}\cdots\uparrow_\ell^{(j_R)}\uparrow\downarrow\cdots}\\
1-2\theta(n\leq j_R)\,,&\text{if }\ket{0;\underline b^{(0)}}=\ket{\cdots\uparrow^{(0)}\bullet\uparrow^{(1)}\cdots\downarrow_\ell\uparrow^{(j_R)}\uparrow\cdots}\\
-1\,,&\text{if }\ket{0;\underline b^{(0)}}=\ket{\cdots\uparrow^{(0)}\bullet\uparrow^{(1)}\cdots\downarrow_\ell\uparrow^{(j_R)}\downarrow\cdots}\,.
\end{cases}
\end{align}
Here, $j_R$ is the index of the particle represented by a spin up at the site $\ell$, or, if the latter is occupied by a spin down, at the site $\ell+1$. The impurity is represented by $\bullet$.
\begin{figure}[ht!]
    %%%%%%%%%%%%%%%%%%%%%%%%%%%%%%%% PROCESS 1
    \hspace{1.5cm}
    \begin{tikzpicture}
    \node[anchor=center] at (0.25,0.75) {$\ell$};
    \filldraw[color=white, fill=orange!60!yellow!90!black, very thick] (-1.0,0) rectangle (-0.5,0.5);
    \filldraw[color=white, fill=orange!60!yellow!90!black, very thick] (-0.5,0) rectangle (0.0,0.5);
    \filldraw[color=white, fill=green!30!blue!50!black, very thick] (0,0) rectangle (0.5,0.5);
    \filldraw [white] (0.25,0.25)  circle (2pt);
    \filldraw[color=white, fill=green!30!blue!50!black, very thick] (0.5,0) rectangle (1.0,0.5);
    \filldraw[color=white, fill=green!30!blue!50!black, very thick] (1.0,0) rectangle (1.5,0.5);
    \node[anchor=west] at (1.75,0.25) {$n=j_R-1$};
    %%%%%
    \filldraw[color=white, fill=orange!60!yellow!90!black, very thick] (0,-0.5) rectangle (0.5,0.0);
    \filldraw[color=white, fill=orange!60!yellow!90!black, very thick] (-0.5,-0.5) rectangle (0.0,0.0);
    \filldraw[color=white, fill=green!30!blue!50!black, very thick] (-1.0,-0.5) rectangle (-0.5,0.0);
    \filldraw [white] (-0.75,-0.25)  circle (2pt);
    \filldraw[color=white, fill=green!30!blue!50!black, very thick] (0.5,-0.5) rectangle (1.0,0.0);
    \filldraw[color=white, fill=green!30!blue!50!black, very thick] (1.0,-0.5) rectangle (1.5,0.0);
    \node[anchor=west] at (1.75,-0.25) {$n=j_R$};
    %%%%%
    \filldraw[color=white, fill=orange!60!yellow!90!black, very thick] (0,-1) rectangle (0.5,-0.5);
    \filldraw[color=white, fill=orange!60!yellow!90!black, very thick] (0.5,-1) rectangle (1.0,-0.5);
    \filldraw[color=white, fill=green!30!blue!50!black, very thick] (-1.0,-1) rectangle (-0.5,-0.5);
    \filldraw [white] (-0.75,-0.75)  circle (2pt);
    \filldraw[color=white, fill=green!30!blue!50!black, very thick] (-0.5,-1) rectangle (0.0,-0.5);
    \filldraw[color=white, fill=green!30!blue!50!black, very thick] (1.0,-1) rectangle (1.5,-0.5);
    \node[anchor=west] at (1.75,-0.75) {$n=j_R+1$};
    %%%%%
    \filldraw[color=white, fill=orange!60!yellow!90!black, very thick] (1.0,-1.5) rectangle (1.5,-1);
    \filldraw[color=white, fill=orange!60!yellow!90!black, very thick] (0.5,-1.5) rectangle (1.0,-1);
    \filldraw[color=white, fill=green!30!blue!50!black, very thick] (-1.0,-1.5) rectangle (-0.5,-1);
    \filldraw [white] (-0.75,-1.25)  circle (2pt);
    \filldraw[color=white, fill=green!30!blue!50!black, very thick] (-0.5,-1.5) rectangle (0.0,-1);
    \filldraw[color=white, fill=green!30!blue!50!black, very thick] (0,-1.5) rectangle (0.5,-1);
    \node[anchor=west] at (1.75,-1.25) {$n=j_R+2$};
    %%%%%
    \draw[color=blue!50!white,rounded corners,very thick] (-0, -1.525) rectangle (0.5,0.525);
    %%%%%
    \node[anchor=west] at (5,-0.5) {$\bra{n;\underline b^{(0)}}\sigma_\ell^z\ket{n;\underline b^{(0)}}=1-2\delta_{n,j_R}-2\delta_{n,j_R+1}$};
    \end{tikzpicture}
    
    %%%%%%%%%%%%%%%%%%%%%%%%%%%%%%%% PROCESS 2
    \hspace{1.5cm}
    \begin{tikzpicture}
    \node[anchor=center] at (0.25,0.75) {$\ell$};
    \filldraw[color=white, fill=orange!60!yellow!90!black, very thick] (-1.0,0) rectangle (-0.5,0.5);
    \filldraw[color=white, fill=orange!60!yellow!90!black, very thick] (-0.5,0) rectangle (0.0,0.5);
    \filldraw[color=white, fill=green!30!blue!50!black, very thick] (0,0) rectangle (0.5,0.5);
    \filldraw [white] (0.25,0.25)  circle (2pt);
    \filldraw[color=white, fill=orange!60!yellow!90!black, very thick] (0.5,0) rectangle (1.0,0.5);
    \filldraw[color=white, fill=green!30!blue!50!black, very thick] (1.0,0) rectangle (1.5,0.5);
    \node[anchor=west] at (1.75,0.25) {$n=j_R-1$};
    %%%%%
    \filldraw[color=white, fill=orange!60!yellow!90!black, very thick] (0,-0.5) rectangle (0.5,0.0);
    \filldraw[color=white, fill=orange!60!yellow!90!black, very thick] (-0.5,-0.5) rectangle (0.0,0.0);
    \filldraw[color=white, fill=green!30!blue!50!black, very thick] (-1.0,-0.5) rectangle (-0.5,0.0);
    \filldraw [white] (-0.75,-0.25)  circle (2pt);
    \filldraw[color=white, fill=orange!60!yellow!90!black, very thick] (0.5,-0.5) rectangle (1.0,0.0);
    \filldraw[color=white, fill=green!30!blue!50!black, very thick] (1.0,-0.5) rectangle (1.5,0.0);
    \node[anchor=west] at (1.75,-0.25) {$n=j_R$};
    %%%%%
    \filldraw[color=white, fill=green!30!blue!50!black, very thick] (0,-1) rectangle (0.5,-0.5);
    \filldraw[color=white, fill=orange!60!yellow!90!black, very thick] (0.5,-1) rectangle (1.0,-0.5);
    \filldraw[color=white, fill=green!30!blue!50!black, very thick] (-1.0,-1) rectangle (-0.5,-0.5);
    \filldraw [white] (-0.75,-0.75)  circle (2pt);
    \filldraw[color=white, fill=orange!60!yellow!90!black, very thick] (-0.5,-1) rectangle (0.0,-0.5);
    \filldraw[color=white, fill=orange!60!yellow!90!black, very thick] (1.0,-1) rectangle (1.5,-0.5);
    \node[anchor=west] at (1.75,-0.75) {$n=j_R+1$};
    %%%%%
    \draw[color=blue!50!white,rounded corners,very thick] (-0, -1.025) rectangle (0.5,0.525);
    %%%%%
    \node[anchor=west] at (5,-0.25) {$\bra{n;\underline b^{(0)}}\sigma_\ell^z\ket{n;\underline b^{(0)}}=1-2\delta_{n,j_R}$};
    \end{tikzpicture}
    
    %%%%%%%%%%%%%%%%%%%%%%%%%%%%%%%% PROCESS 3
    \hspace{1.5cm}
    \begin{tikzpicture}
    \node[anchor=center] at (0.25,0.75) {$\ell$};
    \filldraw[color=white, fill=orange!60!yellow!90!black, very thick] (-1.0,0) rectangle (-0.5,0.5);
    \filldraw[color=white, fill=orange!60!yellow!90!black, very thick] (-0.5,0) rectangle (0.0,0.5);
    \filldraw[color=white, fill=green!30!blue!50!black, very thick] (0,0) rectangle (0.5,0.5);
    \filldraw [white] (0.25,0.25)  circle (2pt);
    \filldraw[color=white, fill=green!30!blue!50!black, very thick] (0.5,0) rectangle (1.0,0.5);
    \filldraw[color=white, fill=orange!60!yellow!90!black, very thick] (1.0,0) rectangle (1.5,0.5);
    \node[anchor=west] at (1.75,0.25) {$n=j_R-1$};
    %%%%%
    \filldraw[color=white, fill=orange!60!yellow!90!black, very thick] (0,-0.5) rectangle (0.5,0.0);
    \filldraw[color=white, fill=orange!60!yellow!90!black, very thick] (-0.5,-0.5) rectangle (0.0,0.0);
    \filldraw[color=white, fill=green!30!blue!50!black, very thick] (-1.0,-0.5) rectangle (-0.5,0.0);
    \filldraw [white] (-0.75,-0.25)  circle (2pt);
    \filldraw[color=white, fill=green!30!blue!50!black, very thick] (0.5,-0.5) rectangle (1.0,0.0);
    \filldraw[color=white, fill=orange!60!yellow!90!black, very thick] (1.0,-0.5) rectangle (1.5,0.0);
    \node[anchor=west] at (1.75,-0.25) {$n=j_R$};
    %%%%%
    \filldraw[color=white, fill=orange!60!yellow!90!black, very thick] (0,-1) rectangle (0.5,-0.5);
    \filldraw[color=white, fill=orange!60!yellow!90!black, very thick] (0.5,-1) rectangle (1.0,-0.5);
    \filldraw[color=white, fill=green!30!blue!50!black, very thick] (-1.0,-1) rectangle (-0.5,-0.5);
    \filldraw [white] (-0.75,-0.75)  circle (2pt);
    \filldraw[color=white, fill=green!30!blue!50!black, very thick] (-0.5,-1) rectangle (0.0,-0.5);
    \filldraw[color=white, fill=orange!60!yellow!90!black, very thick] (1.0,-1) rectangle (1.5,-0.5);
    \node[anchor=west] at (1.75,-0.75) {$n=j_R+1$};
    %%%%%
    \draw[color=blue!50!white,rounded corners,very thick] (-0, -1.025) rectangle (0.5,0.525);
    %%%%%
    \node[anchor=west] at (5,-0.25) {$\bra{n;\underline b^{(0)}}\sigma_\ell^z\ket{n;\underline b^{(0)}}=1-2\theta(n\ge j_R)$};
    \end{tikzpicture}
    
    %%%%%%%%%%%%%%%%%%%%%%%%%%%%%%%% PROCESS 4
    \hspace{1.5cm}
    \begin{tikzpicture}
    \node[anchor=center] at (0.25,0.75) {$\ell$};
    \filldraw[color=white, fill=orange!60!yellow!90!black, very thick] (-1.0,0) rectangle (-0.5,0.5);
    \filldraw[color=white, fill=orange!60!yellow!90!black, very thick] (-0.5,0) rectangle (0.0,0.5);
    \filldraw[color=white, fill=orange!60!yellow!90!black, very thick] (0,0) rectangle (0.5,0.5);
    \filldraw[color=white, fill=green!30!blue!50!black, very thick] (0.5,0) rectangle (1.0,0.5);
    \filldraw [white] (0.75,0.25)  circle (2pt);
    \filldraw[color=white, fill=green!30!blue!50!black, very thick] (1.0,0) rectangle (1.5,0.5);
    \node[anchor=west] at (1.75,0.25) {$n=j_R-1$};
    %%%%%
    \filldraw[color=white, fill=orange!60!yellow!90!black, very thick] (0,-0.5) rectangle (0.5,0.0);
    \filldraw[color=white, fill=green!30!blue!50!black, very thick] (-0.5,-0.5) rectangle (0.0,0.0);
    \filldraw[color=white, fill=orange!60!yellow!90!black, very thick] (-1.0,-0.5) rectangle (-0.5,0.0);
    \filldraw [white] (-0.25,-0.25)  circle (2pt);
    \filldraw[color=white, fill=orange!60!yellow!90!black, very thick] (0.5,-0.5) rectangle (1.0,0.0);
    \filldraw[color=white, fill=green!30!blue!50!black, very thick] (1.0,-0.5) rectangle (1.5,0.0);
    \node[anchor=west] at (1.75,-0.25) {$n=j_R$};
    %%%%%
    \filldraw[color=white, fill=green!30!blue!50!black, very thick] (0,-1) rectangle (0.5,-0.5);
    \filldraw[color=white, fill=orange!60!yellow!90!black, very thick] (0.5,-1) rectangle (1.0,-0.5);
    \filldraw[color=white, fill=orange!60!yellow!90!black, very thick] (-1.0,-1) rectangle (-0.5,-0.5);
    \filldraw[color=white, fill=green!30!blue!50!black, very thick] (-0.5,-1) rectangle (0.0,-0.5);
    \filldraw [white] (-0.25,-0.75)  circle (2pt);
    \filldraw[color=white, fill=orange!60!yellow!90!black, very thick] (1.0,-1) rectangle (1.5,-0.5);
    \node[anchor=west] at (1.75,-0.75) {$n=j_R+1$};
    %%%%%
    \draw[color=blue!50!white,rounded corners,very thick] (-0, -1.025) rectangle (0.5,0.525);
    %%%%%
    \node[anchor=west] at (5,-0.25) {$\bra{n;\underline b^{(0)}}\sigma_\ell^z\ket{n;\underline b^{(0)}}=1-2\theta(n\le j_R)$};
    \end{tikzpicture}
    
    %%%%%%%%%%%%%%%%%%%%%%%%%%%%%%%% PROCESS 5
    \hspace{1.5cm}
    \begin{tikzpicture}
    \node[anchor=center] at (0.25,0.75) {$\ell$};
    \filldraw[color=white, fill=orange!60!yellow!90!black, very thick] (-1.0,0) rectangle (-0.5,0.5);
    \filldraw[color=white, fill=orange!60!yellow!90!black, very thick] (-0.5,0) rectangle (0.0,0.5);
    \filldraw[color=white, fill=orange!60!yellow!90!black, very thick] (0,0) rectangle (0.5,0.5);
    \filldraw[color=white, fill=green!30!blue!50!black, very thick] (0.5,0) rectangle (1.0,0.5);
    \filldraw [white] (0.75,0.25)  circle (2pt);
    \filldraw[color=white, fill=orange!60!yellow!90!black, very thick] (1.0,0) rectangle (1.5,0.5);
    \node[anchor=west] at (1.75,0.25) {$n=j_R-1$};
    %%%%%
    \filldraw[color=white, fill=orange!60!yellow!90!black, very thick] (0,-0.5) rectangle (0.5,0.0);
    \filldraw[color=white, fill=green!30!blue!50!black, very thick] (-0.5,-0.5) rectangle (0.0,0.0);
    \filldraw[color=white, fill=orange!60!yellow!90!black, very thick] (-1.0,-0.5) rectangle (-0.5,0.0);
    \filldraw [white] (-0.25,-0.25)  circle (2pt);
    \filldraw[color=white, fill=orange!60!yellow!90!black, very thick] (0.5,-0.5) rectangle (1.0,0.0);
    \filldraw[color=white, fill=orange!60!yellow!90!black, very thick] (1.0,-0.5) rectangle (1.5,0.0);
    \node[anchor=west] at (1.75,-0.25) {$n=j_R$};
    %%%%%
    \draw[color=blue!50!white,rounded corners,very thick] (-0, -0.525) rectangle (0.5,0.525);
    %%%%%
    \node[anchor=west] at (5,0) {$\bra{n;\underline b^{(0)}}\sigma_\ell^z\ket{n;\underline b^{(0)}}=-1$};
    \end{tikzpicture}
    \caption{The five processes that determine the rules~\eqref{eq:dynamical_rules_app} for the right-moving impurity. Dark blue 
    \protect\tikz{
    \protect\filldraw[color=white,fill=green!30!blue!50!black,very thick] (0,0) rectangle (0.25,0.25);
    }
    and orange 
    \protect\tikz{
    \protect\filldraw[color=white,fill=orange!60!yellow!90!black,very thick] (0,0) rectangle (0.25,0.25);
    } 
    colors represent spins up and down, respectively: two or three consecutive spins down (orange boxes) constitute an impurity, which starts on the left-hand side of the site $\ell$. The white dot represents the spin up corresponding to the particle with index $j_R$. The consequent steps (actions of the local Hamiltonian densities) follow in the downwards direction, and the site $\ell$ is framed.}
    \label{fig:matrix_elements}
\end{figure}

\subsubsection{Decreasing \texorpdfstring{$n$}{TEXT}: the left-moving impurity}

Let us now consider the case of decreasing $n$, in which the impurity passes from the right- to the left-hand side of the site $\ell$. The rules proven in Fig.~\ref{fig:matrix_elements_2} now read
\begin{align}
\label{eq:dynamical_rules_app_2}
\braket{n;\underline b^{(0)}|\sigma_{\ell<-1}^z|n;\underline b^{(0)}}=\begin{cases}
1-2\delta_{n,j_L-1}-2\delta_{n,j_L-2}\,,&\text{if }\ket{0;\underline b^{(0)}}=\ket{\cdots\uparrow\uparrow\uparrow_\ell^{(j_L)}\cdots\uparrow^{(0)}\bullet\uparrow^{(1)}\cdots}\\
1-2\delta_{n,j_L-1}\,,&\text{if }\ket{0;\underline b^{(0)}}=\ket{\cdots\uparrow\downarrow\uparrow_\ell^{(j_L)}\cdots\uparrow^{(0)}\bullet\uparrow^{(1)}\cdots}\\
1-2\theta(n\geq  j_L-1)\,,&\text{if }\ket{0;\underline b^{(0)}}=\ket{\cdots\uparrow\uparrow^{(j_L)}\downarrow_\ell\cdots\uparrow^{(0)}\bullet\uparrow^{(1)}\cdots}\\
1-2\theta(n< j_L)\,,&\text{if }\ket{0;\underline b^{(0)}}=\ket{\cdots\downarrow\uparrow\uparrow^{(j_L)}_\ell\cdots\uparrow^{(0)}\bullet\uparrow^{(1)}\cdots}\\
-1\,,&\text{if }\ket{0;\underline b^{(0)}}=\ket{\cdots\downarrow\uparrow^{(j_L)}\downarrow_\ell\cdots\uparrow^{(0)}\bullet\uparrow^{(1)}\cdots}\,.
\end{cases}
\end{align}
Here, $j_L$ is the index of the particle represented by a spin up at the site $\ell$, or, if the latter is occupied by a spin down, at the site $\ell-1$.
\begin{figure}[ht!]
    %%%%%%%%%%%%%%%%%%%%%%%%%%%%%%%% PROCESS 1
    \hspace{1.5cm}
    \begin{tikzpicture}
    \node[anchor=center] at (0.25,0.75) {$\ell$};
    \filldraw[color=white, fill=green!30!blue!50!black, very thick] (-1.0,0) rectangle (-0.5,0.5);
    \filldraw[color=white, fill=green!30!blue!50!black, very thick] (-0.5,0) rectangle (0.0,0.5);
    \filldraw[color=white, fill=green!30!blue!50!black, very thick] (0,0) rectangle (0.5,0.5);
    \filldraw [white] (0.25,0.25)  circle (2pt);
    \filldraw[color=white, fill=orange!60!yellow!90!black, very thick] (0.5,0) rectangle (1.0,0.5);
    \filldraw[color=white, fill=orange!60!yellow!90!black, very thick] (1.0,0) rectangle (1.5,0.5);
    \node[anchor=west] at (1.75,0.25) {$n=j_L$};
    %%%%%
    \filldraw[color=white, fill=orange!60!yellow!90!black, very thick] (0,-0.5) rectangle (0.5,0.0);
    \filldraw[color=white, fill=green!30!blue!50!black, very thick] (-0.5,-0.5) rectangle (0.0,0.0);
    \filldraw[color=white, fill=green!30!blue!50!black, very thick] (-1.0,-0.5) rectangle (-0.5,0.0);
    \filldraw[color=white, orange!60!yellow!90!black, very thick] (0.5,-0.5) rectangle (1.0,0.0);
    \filldraw[color=white, fill=green!30!blue!50!black, very thick] (1.0,-0.5) rectangle (1.5,0.0);
    \filldraw [white] (1.25,-0.25)  circle (2pt);
    \node[anchor=west] at (1.75,-0.25) {$n=j_L-1$};
    %%%%%
    \filldraw[color=white, fill=orange!60!yellow!90!black, very thick] (0,-1) rectangle (0.5,-0.5);
    \filldraw[color=white, fill=green!30!blue!50!black, very thick] (0.5,-1) rectangle (1.0,-0.5);
    \filldraw[color=white, fill=green!30!blue!50!black, very thick] (-1.0,-1) rectangle (-0.5,-0.5);
    \filldraw[color=white, fill=orange!60!yellow!90!black, very thick] (-0.5,-1) rectangle (0.0,-0.5);
    \filldraw[color=white, fill=green!30!blue!50!black, very thick] (1.0,-1) rectangle (1.5,-0.5);
    \filldraw [white] (1.25,-0.75)  circle (2pt);
    \node[anchor=west] at (1.75,-0.75) {$n=j_L-2$};
    %%%%%
    \filldraw[color=white, fill=green!30!blue!50!black, very thick] (1.0,-1.5) rectangle (1.5,-1);
    \filldraw[color=white, fill=green!30!blue!50!black, very thick] (0.5,-1.5) rectangle (1.0,-1);
    \filldraw[color=white, fill=orange!60!yellow!90!black, very thick] (-1.0,-1.5) rectangle (-0.5,-1);
    \filldraw[color=white, fill=orange!60!yellow!90!black, very thick] (-0.5,-1.5) rectangle (0.0,-1);
    \filldraw[color=white, fill=green!30!blue!50!black, very thick] (0,-1.5) rectangle (0.5,-1);
    \filldraw [white] (1.25,-1.25)  circle (2pt);
    \node[anchor=west] at (1.75,-1.25) {$n=j_L-3$};
    %%%%%
    \draw[color=blue!50!white,rounded corners,very thick] (-0, -1.525) rectangle (0.5,0.525);
    %%%%%
    \node[anchor=west] at (5,-0.5) {$\bra{n;\underline b^{(0)}}\sigma_\ell^z\ket{n;\underline b^{(0)}}=1-2\delta_{n,j_L-1}-2\delta_{n,j_L-2}$};
    \end{tikzpicture}
    
    %%%%%%%%%%%%%%%%%%%%%%%%%%%%%%%% PROCESS 2
    \hspace{1.5cm}
    \begin{tikzpicture}
    \node[anchor=center] at (0.25,0.75) {$\ell$};
    \filldraw[color=white, fill=green!30!blue!50!black, very thick] (-1.0,0) rectangle (-0.5,0.5);
    \filldraw[color=white, fill=orange!60!yellow!90!black, very thick] (-0.5,0) rectangle (0.0,0.5);
    \filldraw[color=white, fill=green!30!blue!50!black, very thick] (0,0) rectangle (0.5,0.5);
    \filldraw [white] (0.25,0.25)  circle (2pt);
    \filldraw[color=white, fill=orange!60!yellow!90!black, very thick] (0.5,0) rectangle (1.0,0.5);
    \filldraw[color=white, fill=orange!60!yellow!90!black, very thick] (1.0,0) rectangle (1.5,0.5);
    \node[anchor=west] at (1.75,0.25) {$n=j_L$};
    %%%%%
    \filldraw[color=white, fill=orange!60!yellow!90!black, very thick] (0,-0.5) rectangle (0.5,0.0);
    \filldraw[color=white, fill=orange!60!yellow!90!black, very thick] (-0.5,-0.5) rectangle (0.0,0.0);
    \filldraw[color=white, fill=green!30!blue!50!black, very thick] (-1.0,-0.5) rectangle (-0.5,0.0);
    \filldraw[color=white, fill=orange!60!yellow!90!black, very thick] (0.5,-0.5) rectangle (1.0,0.0);
    \filldraw[color=white, fill=green!30!blue!50!black, very thick] (1.0,-0.5) rectangle (1.5,0.0);
    \filldraw [white] (1.25,-0.25)  circle (2pt);
    \node[anchor=west] at (1.75,-0.25) {$n=j_L-1$};
    %%%%%
    \filldraw[color=white, fill=green!30!blue!50!black, very thick] (0,-1) rectangle (0.5,-0.5);
    \filldraw[color=white, fill=orange!60!yellow!90!black, very thick] (0.5,-1) rectangle (1.0,-0.5);
    \filldraw[color=white, fill=orange!60!yellow!90!black, very thick] (-1.0,-1) rectangle (-0.5,-0.5);
    \filldraw[color=white, fill=orange!60!yellow!90!black, very thick] (-0.5,-1) rectangle (0.0,-0.5);
    \filldraw[color=white, fill=green!30!blue!50!black, very thick] (1.0,-1) rectangle (1.5,-0.5);
    \filldraw [white] (1.25,-0.75)  circle (2pt);
    \node[anchor=west] at (1.75,-0.75) {$n=j_L-2$};
    %%%%%
    \draw[color=blue!50!white,rounded corners,very thick] (-0, -1.025) rectangle (0.5,0.525);
    %%%%%
    \node[anchor=west] at (5,-0.25) {$\bra{n;\underline b^{(0)}}\sigma_\ell^z\ket{n;\underline b^{(0)}}=1-2\delta_{n,j_L-1}$};
    \end{tikzpicture}
    
    %%%%%%%%%%%%%%%%%%%%%%%%%%%%%%%% PROCESS 3
    \hspace{1.5cm}
    \begin{tikzpicture}
    \node[anchor=center] at (0.25,0.75) {$\ell$};
    \filldraw[color=white, fill=green!30!blue!50!black, very thick] (-1.0,0) rectangle (-0.5,0.5);
    \filldraw[color=white, fill=green!30!blue!50!black, very thick] (-0.5,0) rectangle (0.0,0.5);
    \filldraw[color=white, fill=orange!60!yellow!90!black, very thick] (0,0) rectangle (0.5,0.5);
    \filldraw [white] (-0.25,0.25)  circle (2pt);
    \filldraw[color=white, fill=orange!60!yellow!90!black, very thick] (0.5,0) rectangle (1.0,0.5);
    \filldraw[color=white, fill=orange!60!yellow!90!black, very thick] (1.0,0) rectangle (1.5,0.5);
    \node[anchor=west] at (1.75,0.25) {$n=j_L$};
    %%%%%
    \filldraw[color=white, fill=orange!60!yellow!90!black, very thick] (0,-0.5) rectangle (0.5,0.0);
    \filldraw[color=white, fill=orange!60!yellow!90!black, very thick] (-0.5,-0.5) rectangle (0.0,0.0);
    \filldraw[color=white, fill=green!30!blue!50!black, very thick] (-1.0,-0.5) rectangle (-0.5,0.0);
    \filldraw[color=white, fill=green!30!blue!50!black, very thick] (0.5,-0.5) rectangle (1.0,0.0);
    \filldraw [white] (0.75,-0.25)  circle (2pt);
    \filldraw[color=white, fill=orange!60!yellow!90!black, very thick] (1.0,-0.5) rectangle (1.5,0.0);
    \node[anchor=west] at (1.75,-0.25) {$n=j_L-1$};
    %%%%%
    \filldraw[color=white, fill=green!30!blue!50!black, very thick] (0,-1) rectangle (0.5,-0.5);
    \filldraw[color=white, fill=green!30!blue!50!black, very thick] (0.5,-1) rectangle (1.0,-0.5);
    \filldraw[color=white, fill=orange!60!yellow!90!black, very thick] (-1.0,-1) rectangle (-0.5,-0.5);
    \filldraw[color=white, fill=orange!60!yellow!90!black, very thick] (-0.5,-1) rectangle (0.0,-0.5);
    \filldraw[color=white, fill=orange!60!yellow!90!black, very thick] (1.0,-1) rectangle (1.5,-0.5);
    \filldraw [white] (0.75,-0.75)  circle (2pt);
    \node[anchor=west] at (1.75,-0.75) {$n=j_L-2$};
    %%%%%
    \draw[color=blue!50!white,rounded corners,very thick] (-0, -1.025) rectangle (0.5,0.525);
    %%%%%
    \node[anchor=west] at (5,-0.25) {$\bra{n;\underline b^{(0)}}\sigma_\ell^z\ket{n;\underline b^{(0)}}=1-2\theta(n\ge j_L-1)$};
    \end{tikzpicture}
    
    %%%%%%%%%%%%%%%%%%%%%%%%%%%%%%%% PROCESS 4
    \hspace{1.5cm}
    \begin{tikzpicture}
    \node[anchor=center] at (0.25,0.75) {$\ell$};
    \filldraw[color=white, fill=orange!60!yellow!90!black, very thick] (-1.0,0) rectangle (-0.5,0.5);
    \filldraw[color=white, fill=green!30!blue!50!black, very thick] (-0.5,0) rectangle (0.0,0.5);
    \filldraw[color=white, fill=green!30!blue!50!black, very thick] (0,0) rectangle (0.5,0.5);
    \filldraw[color=white, fill=orange!60!yellow!90!black, very thick] (0.5,0) rectangle (1.0,0.5);
    \filldraw [white] (0.25,0.25)  circle (2pt);
    \filldraw[color=white, fill=orange!60!yellow!90!black, very thick] (1.0,0) rectangle (1.5,0.5);
    \node[anchor=west] at (1.75,0.25) {$n=j_L$};
    %%%%%
    \filldraw[color=white, fill=orange!60!yellow!90!black, very thick] (0,-0.5) rectangle (0.5,0.0);
    \filldraw[color=white, fill=green!30!blue!50!black, very thick] (-0.5,-0.5) rectangle (0.0,0.0);
    \filldraw[color=white, fill=orange!60!yellow!90!black, very thick] (-1.0,-0.5) rectangle (-0.5,0.0);
    \filldraw[color=white, fill=orange!60!yellow!90!black, very thick] (0.5,-0.5) rectangle (1.0,0.0);
    \filldraw[color=white, fill=green!30!blue!50!black, very thick] (1.0,-0.5) rectangle (1.5,0.0);
    \filldraw [white] (1.25,-0.25)  circle (2pt);
    \node[anchor=west] at (1.75,-0.25) {$n=j_L-1$};
    %%%%%
    \filldraw[color=white, fill=orange!60!yellow!90!black, very thick] (0,-1) rectangle (0.5,-0.5);
    \filldraw[color=white, fill=green!30!blue!50!black, very thick] (0.5,-1) rectangle (1.0,-0.5);
    \filldraw[color=white, fill=orange!60!yellow!90!black, very thick] (-1.0,-1) rectangle (-0.5,-0.5);
    \filldraw[color=white, fill=orange!60!yellow!90!black, very thick] (-0.5,-1) rectangle (0.0,-0.5);
    \filldraw[color=white, fill=green!30!blue!50!black, very thick] (1.0,-1) rectangle (1.5,-0.5);
    \filldraw [white] (1.25,-0.25)  circle (2pt);
    \node[anchor=west] at (1.75,-0.75) {$n=j_L-2$};
    %%%%%
    \draw[color=blue!50!white,rounded corners,very thick] (-0, -1.025) rectangle (0.5,0.525);
    %%%%%
    \node[anchor=west] at (5,-0.25) {$\bra{n;\underline b^{(0)}}\sigma_\ell^z\ket{n;\underline b^{(0)}}=1-2\theta(n<j_L)$};
    \end{tikzpicture}
    
    %%%%%%%%%%%%%%%%%%%%%%%%%%%%%%%% PROCESS 5
    \hspace{1.5cm}
    \begin{tikzpicture}
    \node[anchor=center] at (0.25,0.75) {$\ell$};
    \filldraw[color=white, fill=orange!60!yellow!90!black, very thick] (-1.0,0) rectangle (-0.5,0.5);
    \filldraw[color=white, fill=green!30!blue!50!black, very thick] (-0.5,0) rectangle (0.0,0.5);
    \filldraw[color=white, fill=orange!60!yellow!90!black, very thick] (0,0) rectangle (0.5,0.5);
    \filldraw[color=white, fill=orange!60!yellow!90!black, very thick] (0.5,0) rectangle (1.0,0.5);
    \filldraw [white] (-0.25,0.25)  circle (2pt);
    \filldraw[color=white, fill=orange!60!yellow!90!black, very thick] (1.0,0) rectangle (1.5,0.5);
    \node[anchor=west] at (1.75,0.25) {$n=j_L$};
    %%%%%
    \filldraw[color=white, fill=orange!60!yellow!90!black, very thick] (0,-0.5) rectangle (0.5,0.0);
    \filldraw[color=white, fill=orange!60!yellow!90!black, very thick] (-0.5,-0.5) rectangle (0.0,0.0);
    \filldraw[color=white, fill=orange!60!yellow!90!black, very thick] (-1.0,-0.5) rectangle (-0.5,0.0);
    \filldraw[color=white, fill=green!30!blue!50!black, very thick] (0.5,-0.5) rectangle (1.0,0.0);
    \filldraw [white] (0.75,-0.25)  circle (2pt);
    \filldraw[color=white, fill=orange!60!yellow!90!black, very thick] (1.0,-0.5) rectangle (1.5,0.0);
    \node[anchor=west] at (1.75,-0.25) {$n=j_L-1$};
    %%%%%
    \draw[color=blue!50!white,rounded corners,very thick] (-0, -0.525) rectangle (0.5,0.525);
    %%%%%
    \node[anchor=west] at (5,0) {$\bra{n;\underline b^{(0)}}\sigma_\ell^z\ket{n;\underline b^{(0)}}=-1$};
    \end{tikzpicture}
    \caption{The five processes that determine the rules~\eqref{eq:dynamical_rules_app_2} for the left-moving impurity. Dark blue 
    \protect\tikz{
    \protect\filldraw[color=white,fill=green!30!blue!50!black,very thick] (0,0) rectangle (0.25,0.25);
    }
    and orange 
    \protect\tikz{
    \protect\filldraw[color=white,fill=orange!60!yellow!90!black,very thick] (0,0) rectangle (0.25,0.25);
    } 
    colors represent spins up and down, respectively: two or three consecutive spins down (orange boxes) constitute an impurity, which now starts on the right-hand side of the site $\ell$. The white dot represents the spin up corresponding to the particle with index $j_L$. The consequent steps (actions of the local Hamiltonian densities) follow in the downwards direction, and the site $\ell$ is framed.}
    \label{fig:matrix_elements_2}
\end{figure}

\subsection{Asymptotic profiles of magnetisation}
\label{app:profiles_magnetisation}

In the previous section we have established the rules for the computation of the matrix elements of $\sigma_\ell^z$. With this let us return to its expectation value~\eqref{eq:bessel_formula}. For the asymptotic profiles only large values of $n$ are relevant, whence the rules~\eqref{eq:dynamical_rules_app} and~\eqref{eq:dynamical_rules_app_2} reduce to
\begin{align}
\label{eq:asymptotic_rules_app}
\braket{n;\underline b^{(0)}|\sigma_\ell^z|n;\underline b^{(0)}}\sim\begin{cases}
1\,,&\text{for }\ket{\cdots\bullet\cdots\uparrow_\ell\updownarrow\uparrow\cdots}\vee\ket{\cdots\uparrow\updownarrow\uparrow_\ell\cdots\bullet\cdots}\\
1-2\theta(n\geq x(\ell))\,,&\text{for }\ket{\cdots\bullet\cdots\uparrow_\ell\uparrow\downarrow\cdots}\vee\ket{\cdots\uparrow\uparrow\downarrow_\ell\cdots\bullet\cdots}\\
1-2\theta(n\leq x(\ell))\,,&\text{for }\ket{\cdots\bullet\cdots\downarrow_\ell\uparrow\uparrow\cdots}\vee\ket{\cdots\downarrow\uparrow\uparrow_\ell\cdots\bullet\cdots}\\
-1\,,&\text{for }\ket{\cdots\bullet\cdots\downarrow_\ell\uparrow\downarrow\cdots}\vee\ket{\cdots\downarrow\uparrow\downarrow_\ell\cdots\bullet\cdots}\,.
\end{cases}
\end{align}
Here, $\updownarrow$ stands for either a spin up or a spin down, while $x(\ell)$
corresponds to the index of the particle represented by the spin up at the site $\ell$ (or the neighbouring site). Asymptotically we compute it by observing $\ell'_{x(\ell)}\sim \ell/2$ and using Eq.~\eqref{eq:mapping1}. For example, for large $x(\ell)>0$ one has
\begin{align}
    \ell'_{x(\ell)} = x(\ell)-\theta(x(\ell)\leq n)-\sum^{x(\ell)-1}_{m=1}b^{(0)}_{m}(1-b^{(0)}_{m+1})\sim\frac{\ell}{2}\, ,
\end{align}
where any terms of magnitude $O(1)$ can be neglected. For large $|\ell|$ we can thus compute $x(\ell)$ from the integral equation
\begin{align}
    \frac{1}{\ell}\int_{0}^{x(\ell)}{\rm d}n \, \xi(n)=\frac{1}{2}\, ,
\end{align}
where $\xi(n)$ is the average $1-b^{(0)}_j(1-b^{(0)}_{j+1})$ around the particle with index $n$.

Using the asymptotic rules~\eqref{eq:asymptotic_rules_app} we now readily obtain Eq.~\eqref{eq:asymptotic_magnetisation}. Consider, for example, the second rule in Eq.~\eqref{eq:asymptotic_rules_app}:
\begin{align}
\begin{gathered}
\label{eq:magnetization_exact}
\sum_{n}\bra{n;\underline{b}^{(0)}}\sigma_\ell^z\ket{n;\underline{b}^{(0)}} |J_{n}(4 \mathcal{J} t)|^2=1-2\sum_n\theta(n\ge x(\ell))|J_n(4\mathcal{J}t)|^2=\\
=1-2\!\int\!\frac{{\rm d}k}{2\pi}\!\int\!\frac{{\rm d}p}{2\pi}\!\sum_{n\ge x(\ell)}\! e^{-i (k-p)n-i4\mathcal{J}t(\cos k-\cos p)}
=1-2\!\int\!\frac{{\rm d}k}{2\pi}\!\int\!\frac{{\rm d}p}{2\pi}\frac{e^{-i (k-p)x(\ell)-i4\mathcal{J}t(\cos k-\cos p)}}{1-e^{-i(k-p)}}\sim\\
\sim 1-\frac{1}{\pi}\int{\rm d}p\,\theta\Big(\sin p-\frac{x(\ell)}{4\mathcal{J}t}\Big)=\frac{2}{\pi}\arcsin\Big(\frac{x(\ell)}{4\mathcal{J}t}\Big)\,.
\end{gathered}
\end{align}

\section{Correlation functions in the weakly-interacting case}
\label{app:special_protocol}

\subsection{The idea behind the computation}
\label{app:special_protocol_SpinProfiles}

In this section we show the idea behind the exact computation of any correlation function for the weakly-interacting scenario discussed in the main text, in which the unperturbed state at time $t=0^-$ is
\begin{equation}
    \ket{\Psi(0^-)}=\ket{\cdots\downarrow\uparrow\downarrow\uparrow \underset{m'}{\underline{\uparrow\uparrow}}\uparrow\uparrow \cdots \uparrow\uparrow \underset{m'+M-1}{\underline{\uparrow\uparrow}} \downarrow\uparrow\downarrow\uparrow\cdots}\,.
\end{equation}
Essentially, the sums over $n_1$ and $n_2$, e.g., in Eq.~\eqref{eq:off_diagonal}, are split in such a way that the indices referring to eigenstates with the same characteristics are grouped together. In particular, 
we can distinguish four different classes of eigenstates $\ket{n;\underline{b}^{(0)}}$:
\begin{itemize}
    \item $n<m'$: the macrosite $n$ is in the configuration $\downarrow \downarrow$, the macrosites $m',m'+1,...,m'+M-1$ are each in the configuration $\uparrow \uparrow$, all the other macrosites are in the configuration $\downarrow\uparrow$;
    \item $n=m'+2a,a\in\{0,1,...,M-1\}$: the macrosite $m'+a-1$ is in the configuration $\uparrow \downarrow$, the macrosites $m'-1,m',...,m'+a-2,m'+a+1,...,m'+M-1$ are each in the configuration $\uparrow \uparrow$ (note that the latter set is empty for $M=1$), all the other macrosites are in the configuration $\downarrow\uparrow$;
    \item $n=m'+2b+1,b\in\{0,1,...,M-2\}$: (note that this set is empty for $M=1$) the macrosite $m'+b$ is in the configuration $\downarrow \downarrow$, the macrosites $m'-1,m',...,m'+b-1,m'+b+1,...,m'+M-1$ are each in the configuration $\uparrow \uparrow$, all the other macrosites are in the configuration $\downarrow\uparrow$;
    \item $n\ge m'+2M-1$: the macrosite $n-M$ is in the configuration $\downarrow \downarrow$, the macrosites $m'-1,m',...,m'+M-2$ are each in the configuration $\uparrow \uparrow$, all the other macrosites are in the configuration $\downarrow\uparrow$.
\end{itemize}
Using Eq.~\eqref{eq:bessel_formula}, we obtain
\begin{align}
\begin{aligned}
    \braket{\sigma^z_{2\ell'-1}}_t=&
    \sum_{a=0}^{M-2} 
    |J_{m'+2a}(4 \mathcal{J} t)|^2
    \left(
    2\chi_{[m'-1,m'+a-1]}(\ell')+2\chi_{[m'+a+1,m'+M-1]}(\ell') -1
    \right)
     \\
     &+\sum_{b=0}^{M-2} 
    |J_{m'+2b+1}(4\mathcal{J} t)|^2
    \left(
    2\chi_{[m'-1,m'+b-1]}(\ell')+2\chi_{[m'+b+1,m'+M-1]}(\ell') -1
    \right)\\
    &+\sum_{n=-\infty}^{m'-1} 
    |J_{n}(4 \mathcal{J} t)|^2
    \left(
    2\chi_{[m',m'+M-1]}(\ell') -1
    \right)
    \\
    &+\sum_{n=m'+2M-2}^{\infty} 
    |J_{n}(4\mathcal{J} t)|^2
    \left(
    2\chi_{[m'-1,m'+M-2]}(\ell') -1
    \right)
%     \\
%     =2\delta_{\ell',m'+M-1} \sum_{n=-\infty}^{m'-1} 
%     |J_{ -n}(4 J t)|^2
%     + 2\delta_{\ell',m'-1} \sum_{n=m'+2M-2}^{+\infty} 
%     |J_{ -n}(4 J t)|^2
%   \\ +2(\delta_{\ell', m'+M-1}+\delta_{\ell',m'-1})\sum_{n=m'}^{m'+2M-3} 
%     |J_{ -n}(4 J t)|^2 -1
%     \\
%     +2\theta(m'\le \ell'\le m'+M-2) \sum_{n=-\infty}^{m'-1} 
%     |J_{ -n}(4 J t)|^2
%     + 2\theta(m'\le \ell'\le m'+M-2) \sum_{n=m'+2M-2}^{+\infty} 
%     |J_{ -n}(4 J t)|^2
%     \\+2\theta(m'\le \ell'\le m'+M-2)\left(
%     \sum_{n=m'}^{m'+2M-3} 
%     |J_{ -n}(4 J t)|^2
%     -
%     \left(|J_{ +m'-2\ell'-1}(4 J t)|^2+|J_{ +m'-2\ell'}(4 J t)|^2\right)
%     \right)
\, ,
\end{aligned}
\end{align}
for the local magnetisation at the odd sites. Here, $\chi_{[a,b]}(x)$ is the characteristic function on the interval $[a,b]$, i.e., it is equal to one for $x\in[a,b]$ and zero otherwise.
%where we used $\sum_{n=-\infty}^{+\infty}J_n^2(x)=1,\forall x$.
A straightforward calculation now yields
\begin{equation}
    \braket{\sigma^z_{2\ell'-1}}_t=
    \left\{
    \begin{aligned}
    &2f(m'+2M-2,t)-1\,, && \text{if } \ell'=m'+M-1
    \\
    &1-2f(m',t), &&\text{if } \ell'=m'-1
    \\
    &1-2|J_{m'-2\ell'-1}(4 \mathcal{J} t)|^2-2|J_{m'-2\ell'}(4 \mathcal{J} t)|^2\,, 
    &&\text{if } \ell'\in\{m',m'+1,...,m'+M-2\}
    \\
    &-1\,, &&\text{otherwise}\,,
    \end{aligned}
    \right. \,
\end{equation}
where $f(x,t)$ is defined in Eq.~\eqref{eq:f_function}.

Let us now consider an example of a non-diagonal operator, and compute it using Eq.~\eqref{eq:off_diagonal}.
The operators $\sigma^x$ and $\sigma^y$ have zero expectation value, since the states $\bra{n_1;\underline{b}^{(0)}}$ and $\ket{n_2;\underline{b}^{(0)}}$ in Eq.~\eqref{eq:off_diagonal} contain the same number of spins up.
To give an example of a non-trivial non-diagonal operator, we need to consider correlation functions involving an even number of $\sigma^x$ and $\sigma^y$, such as $\braket{\sigma^x_{2\ell'-2d}\sigma^x_{2\ell'}}$, for $d\in\mathbb{Z}^+$.
In order to get a non-zero contribution from a generic term of the sum, the overall action of the spin-flips on $\ket{n_2;\underline{b}^{(0)}}$ should not change the background.
By dividing the sum over $n_2$ in the sets that we introduced above, the only options are
\begin{multline}
    \sigma^x_{2\ell'-2d}\sigma^x_{2\ell'}\ket{n_2;\underline{b}^{(0)}}\rightarrow\\
    \left\{
    \begin{aligned}
    &\delta_{n_2,\ell'-d}\ket{\ell';\underline{b}^{(0)}} + \delta_{n_2,\ell'}\chi_{(-\infty,m')}(\ell')\ket{\ell'-d;\underline{b}^{(0)}}\,,
    && \text{if }n_2<m'
    \\
    &\delta_{n_2\!-\!M,\ell'\!-\!d}\ket{\ell'\!+\!M\!;\!\underline{b}^{(0)}}
    \!+\!\delta_{n_2\!-\!M,\ell'}\chi_{[m'\!+\!d\!+\!M\!-\!2,\infty)}(\ell')\ket{\ell'  \!+\!M\!-\!d;\underline{b}^{(0)}} ,
    &&\text{if } n_2\ge m'+2M-2 
    \\
    &\delta_{d,1}\delta_{\ell',m'+a} \ket{m'+2a+1;\underline{b}^{(0)}}\,, && \text{if }n_2=m'+2a
    \\
    &\delta_{d,1}\delta_{\ell',m'+a} \ket{m'+2a;\underline{b}^{(0)}}, && \text{if }n_2=m'+2a+1\,,
    \end{aligned}
    \right. \,
\end{multline}
where $a\in\{1,2,...,M-2\}$ and we reported only the cases in which the background is unchanged (whence we used ``$\rightarrow$'' instead of ``$=$'').
By inserting back such result into the initial expression, we finally have
\begin{multline}
\braket{\sigma^x_{2(\ell'-d)}\sigma^x_{2\ell'} }_t
    =
    2\cos\bigl(\frac{\pi}{2}d\bigr)
    \Bigl(
    \theta(\ell'<m') 
    J_{ \ell'}(4 \mathcal{J}t)J_{ \ell'-d}(4\mathcal{J}t)
    +\\+
    \theta(\ell'\ge m'+M+d-2) 
    J_{\ell'+M}(4 \mathcal{J} t)J_{\ell'+M-d}(4 \mathcal{J}t)
    \Bigr)
\,.
\end{multline}
All the other one- and two-point functions are obtained analogously.

\subsection{List of the two-point correlation functions}
\be
\ba
\braket{\sigma^\alpha_{2\ell'-j} \sigma^\beta_{2m'-1+j}}=&0\,,\\
\braket{\sigma^\alpha_{2(\ell'-d)-1}\sigma^\alpha_{2\ell'-1}}=&0\,,\\
\braket{\sigma^\alpha_{2(\ell'-d)}\sigma^\beta_{2\ell'} }
    =&
    2\cos\Bigl(\frac{\pi}{2}(\beta-\alpha+d)\Bigr)
    [
    \theta(\ell'<m') 
    J_{\ell'}(4 J t)J_{\ell'-d}(4 \mathcal{J}t)
    \\&\qquad+
    \theta(\ell'\ge m'+M+d-2) 
    J_{\ell'+M}(4 J t)J_{\ell'+M-d}(4 \mathcal{J}t)
    \\&\qquad+
    \delta_{d,1}\theta(m'\le \ell' \le m'+M-2)
    J_{2\ell'-m'+1}(4 \mathcal{J} t)J_{2\ell'-m'}(4 \mathcal{J} t)
    ]\,,\\
\braket{\sigma^{3-\alpha}_{2(\ell'-d)-1}\sigma^{\alpha}_{2\ell'-1} }=&       2(-1)^\alpha\delta_{d,1}\theta(m'\le \ell'\le m'+M-1)  J_{m'-2\ell'+1}(4      \mathcal{J}t)J_{m'-2\ell'}(4\mathcal{J} t)\,,\\
\braket{\sigma^z_{2(\ell'-d)-1}\sigma^z_{2\ell'-1} } =&
    1-2\delta_{m'+M-1,\ell'-d}f(m'+2M-2,t)
    \\&+2\delta_{m'-1,\ell'-d}(-1+f(m',t)+\theta(d<M)(2f(m',t)-1))
    \\&-2\delta_{\ell',m'+M-1}(f(m'+2M-2,t)+2\theta(d<M)f(m'+2M-2,t))
    \\&+2\delta_{\ell',m'-1}(-1+f(m',t))
    +4\delta_{d,M}\delta_{m'-1,\ell'-d}(f(m'+2M-2,t)-f(m',t))
    \\&-2\theta(d\ge M)(\theta(m'\le \ell' < m'+M-1)+\theta(m'+d\le \ell'<m'+d+M-1))
    \\&-2\theta(d<M)(\theta(m'\le \ell'<m'+d-1)+\theta(m'+M-1\le \ell'<m'+d+M-1))
    \\&+2(\theta(m'\le \ell'-d\le m'+M-2)-2\theta(m'+d\le \ell'\le m'+M-1))\\
    &\qquad\times(J^2_{m'-2\ell'+2d}(4 \mathcal{J} t)+J^2_{m'-2\ell'+2d-1}(4 J t))
    \\&+2(\theta(m'\le \ell'\le m'+M-2)-2\theta(m'+d-1\le \ell'\le m'+M-2))\\
    &\qquad\times(J^2_{m'-2\ell'}(4 \mathcal{J} t)+J^2_{m'-2\ell'-1}(4 \mathcal{J} t))\,,\\
\braket{\sigma^z_{2(\ell'-d)}\sigma^z_{2\ell'} } =&
    1-2\theta(\ell'-d<m')J^2_{\ell'-d}(4 \mathcal{J} t)
    -2\theta(\ell'<m')J^2_{\ell'}(4 \mathcal{J} t)
    \\
    &-2\theta(\ell'-d\ge m'+M-2)J^2_{\ell'+M-d}(4\mathcal{J} t)
    -2\theta(\ell' \ge m'+M-2)J^2_{\ell'+M}(4 \mathcal{J} t)
    \\
    &-2\theta(m'-1\le \ell'-d < m'+M-2)J^2_{m'-2\ell'+2d-2}(4 \mathcal{J} t)\\
    &-2\theta(m'-1\le \ell' < m'+M-2)J^2_{m'-2\ell'-2}(4 \mathcal{J} t)
    \\
    &-2\theta(m'\le \ell'-d < m'+M-1)J^2_{m'-2\ell'+2d-1}(4 \mathcal{J} t)\\
    &-2\theta(m'\le \ell' < m'+M-1)J^2_{m'-2\ell'-1}(4 \mathcal{J} t)\,,\\
\braket{\sigma^z_{2a'}\sigma^z_{2\ell'-1} } =&
    -1 + 2\theta(a'<m')J^2_{a'}(4 \mathcal{J}t) + 2\theta(a'\ge m'+M-2)J^2_{a'+M}(4 \mathcal{J} t) \\
    &+ 2\theta(m'-1\le a' \le m'+M-3)J^2_{m'-2a'-2}(4 \mathcal{J} t) \\
    &+ 2\theta(m'\le a' \le m'+M-2)J^2_{m'-2a'-1}(4 \mathcal{J} t)
    \\&+
    2\delta_{\ell',m'-1}[
    1 - f(m',t) - 2\theta(a'\ge m'+M-2)J^2_{a'+M}(4 \mathcal{J} t)
    ]
    \\&+
    2\delta_{\ell',m'+M-1}[
    f(m'+2M-2,t) - 2\theta(a'<m')J^2_{a'}(4 \mathcal{J}t)
    ]
    \\&+
    2\theta(m'\le \ell' \le m'+M-2)[
    1 - J^2_{m'-2\ell'}(4 \mathcal{J} t) - J^2_{m'-2\ell'-1}(4 \mathcal{J} t)\\&\qquad
    -2\theta(a'<m')J^2_{a'}(4 \mathcal{J} t)
    -2\theta(a'\ge m'+M-2)J^2_{a'+M}(4 \mathcal{J}t)
    \\&\qquad+ 2\delta_{\ell',a'+1}J^2_{m'-2\ell'}(4 \mathcal{J} t) + 2\delta_{a',\ell'}J^2_{m'-2a'-1}(4 \mathcal{J} t)
    ]
    \\&-
    4\theta(m'-1\le \ell' \le m'+M-1)[
   \theta(m'-1\le a' \le m'+M-3)J^2_{m'-2a'-2}(4 \mathcal{J} t) 
   \\&\qquad+ \theta(m'\le a' \le m'+M-2)J^2_{m'-2a'-1}(4 \mathcal{J}t) 
    ]\,,
\ea
\ee
where $\alpha,\beta\in\{1,2\}$, $\sigma^1=\sigma^x$, $\sigma^2=\sigma^y$, $j\in\{0,1\}$, and $d\geq 1$.

\section{Details on numerical simulations}
\label{app:numerics}
%%%
\begin{figure}[h!]
  %\hspace{-1.5em}
  \centering
  \includegraphics[width=0.65\textwidth]{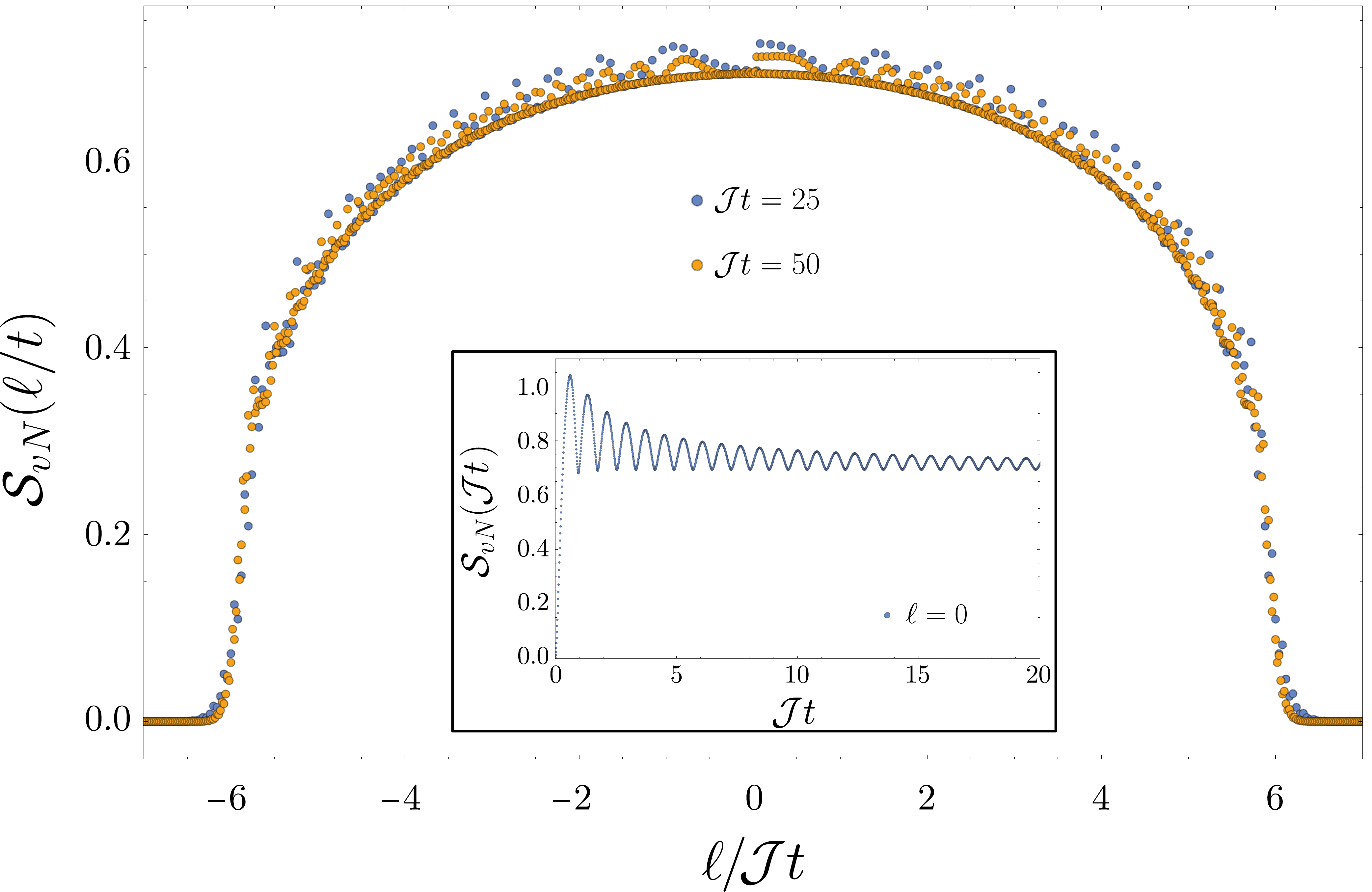}
  \caption{Bipartite entanglement entropy $S^{vN}_{\ell}(t)$ for $t=25$ (blue points) and $t=50$ (orange points). The inhomogeneity in the initial state $\ket{\cdots\uparrow\uparrow\downarrow \uparrow\uparrow\downarrow \uparrow\downarrow\downarrow  \uparrow\uparrow\downarrow \uparrow\uparrow\downarrow\cdots}$ spreads with time and forms a nontrivial ballistic profile. The inset shows the entanglement entropy $S^{vN}_0(t)$ in the center of the system. Oscillations vanish with time, while the entanglement reaches its limiting value.}
  \label{fig:Entanglement_profile}
\end{figure}
%%%
Numerical calculations are performed using C++ Itensor library~\cite{itensor}.
We use the time evolving block decimation (TEBD) as a time-evolution scheme, where $e^{-i \delta t H}$ for the folded model is approximated by Trotter-Suzuki gates~\cite{trotter, trotter2, trotter3, folded:2}. In order to obtain higher approximation, we represent the Hamiltonian~\eqref{eq:Hamiltonian} as a sum 
\begin{align}
H=\sum_{\ell} h_{3\ell,3\ell+1,3\ell+2} +\sum_{\ell} h_{3\ell+1,3\ell+2,3\ell+3} + \sum_{\ell}h_{3\ell+2,3\ell+3,3\ell+4} = H_1 + H_2 + H_3\, ,
\end{align}
with commuting terms within each $H_n$. The time evolution operator reads
\begin{equation}
    e^{-i\delta t (H_1+ H_2+ H_3)} = e^{- i\frac{\delta t}{2} H_1  }e^{-i\frac{\delta t}{2} H_2}e^{-i\delta t H_3}e^{-i\frac{\delta t}{2} H_2}e^{-i\frac{\delta t}{2}H_1} + \mathcal{O}(\delta t^3)\, ,
\end{equation}
where the time step is chosen as $\delta t = 0.01\mathcal{J}^{-1}$. 
Numerical simulations for the XXZ model are performed using the matrix-product-based approximation to $e^{-i\delta t H}$ of the third order, denoted by $W^{\rm II}$ in Ref~\cite{zaletel, bidzhiev}. As a time step in this case we chose $\delta t = 0.001\mathcal{J}^{-1}$ to reduce the potential errors from the large-anisotropy term.

The level of matrix-product state truncation is determined by some maximal value of the discarded weight, which we chose to be $\delta \lambda =10^{-12}$. It turns out that for the most of the calculations in the folded XXZ model the maximal bond dimension is $\chi_{\rm max}=3$ during the whole time evolution, meanwhile the bipartite entanglement entropy $S_{vN}$ forms a nontrivial scaling profile in the ballisitc limit -- see Fig.~\ref{fig:Entanglement_profile}.

\end{appendix}

\end{document}